\documentclass[a4paper,11pt,preprintnumbers]{article}
\pdfoutput=1

\usepackage{amssymb,mathrsfs,amsmath}
\usepackage{a4wide}
\usepackage{color,xcolor}
\usepackage{slashed,soul}
\usepackage{graphicx}
\usepackage{amsfonts}
\usepackage{lscape}
\def\linkcolor{cyan!70!black}

\usepackage[
colorlinks=true
,urlcolor=\linkcolor
,anchorcolor=\linkcolor
,citecolor=\linkcolor
,filecolor=\linkcolor
,linkcolor=\linkcolor
,menucolor=\linkcolor
,linktocpage=true
,pdfproducer=medialab
,pdfa=true
]{hyperref}
\usepackage{amsthm}
\usepackage{booktabs}
\usepackage{array}
\usepackage{rotating}
\usepackage[numbers, sort&compress]{natbib}
\usepackage{multirow}
\usepackage{float}
\usepackage[utf8]{inputenc}
\usepackage[T1]{fontenc}
\usepackage{appendix}
\usepackage{epsfig}
\usepackage{orcidlink}
\usepackage{comment}
\usepackage{adjustbox}

\newcommand{\beq}{\begin{equation}} 
\newcommand{\eeq}{\end{equation}} 
\newcommand{\ba}{\begin{array}}  
\newcommand{\ea}{\end{array}} 
\newcommand{\bea}{\begin{eqnarray}}  
\newcommand{\eea}{\end{eqnarray} }  
\newcommand{\bal}{\begin{align}}
\newcommand{\eal}{\end{align}}   
\newcommand{\bi}{\begin{itemize}}  
\newcommand{\ei}{\end{itemize}}  
\newcommand{\ben}{\begin{enumerate}}  
\newcommand{\een}{\end{enumerate}}  
\newcommand{\bc}{\begin{center}}
\newcommand{\ec}{\end{center}} 
\newcommand{\bt}{\begin{table}}
\newcommand{\et}{\end{table}}  
\newcommand{\btb}{\begin{tabular}}
\newcommand{\etb}{\end{tabular}}

\newcommand{\hc}{\mathrm{h.c.}}
\renewcommand{\O}{\mathcal{O}}

\newcommand{\opd}[2]{\O^{#2}_{#1}}
\newcommand{\wcL}[2]{c^{#2}_{#1}} 

\def\arrvline{\hfil\kern\arraycolsep\vline\kern-\arraycolsep\hfilneg}
\newcommand{\sw}{s_{\mathrm{w}}}

\newcommand{\nsi}[2]{\varepsilon^{#1}_{#2}}
\newcommand{\cevns}{CE$\nu$NS}

\allowdisplaybreaks

\let\OLDthebibliography\thebibliography
\renewcommand\thebibliography[1]{
  \OLDthebibliography{#1}
  \setlength{\parskip}{0pt}
  \setlength{\itemsep}{0pt plus 0.3ex}
}

\begin{document}

\vspace{1cm}

\begin{titlepage}

\begin{flushright}
IFT-UAM/CSIC-24-151\\
FTUV-24-1025.8856\\
 \end{flushright}
\vspace{0.2truecm}

\begin{center}
\renewcommand{\baselinestretch}{1.8}\normalsize
\boldmath
{\LARGE\textbf{
Improving the Global SMEFT Picture \\ with Bounds on Neutrino NSI
}}
\unboldmath
\end{center}

\vspace{0.4truecm}

\renewcommand*{\thefootnote}{\fnsymbol{footnote}}

\begin{center}

{
Pilar Coloma$^1$\footnote{\href{mailto:pilar.coloma@ift.csic.es}{pilar.coloma@ift.csic.es}}\orcidlink{0000-0002-1164-9900},
Enrique Fern\'andez-Mart\'inez$^1$\footnote{\href{mailto:enrique.fernandez@csic.es}{enrique.fernandez@csic.es}}\orcidlink{0000-0002-6274-4473}, 
Jacobo L\'opez-Pav\'on$^2$\footnote{\href{mailto:Jacobo.Lopez@uv.es}{jacobo.lopez@uv.es}}\orcidlink{0000-0002-9554-5075},
Xabier~Marcano$^1$\footnote{\href{mailto:xabier.marcano@uam.es}{xabier.marcano@uam.es}}\orcidlink{0000-0003-0033-0504},
Daniel Naredo-Tuero$^1$\footnote{\href{mailto:daniel.naredo@uam.es}{daniel.naredo@ift.csic.es}}\orcidlink{0000-0002-5161-5895}
}
and Salvador Urrea$^{2,3}$\footnote{\href{mailto:salvador.urrea@ific.uv.es}{salvador.urrea@ijclab.in2p3.fr}}\orcidlink{0000-0002-7670-232X}

\vspace{0.7truecm}

{\footnotesize
$^1$Instituto de F\'{\i}sica Te\'orica UAM/CSIC,\\
Universidad Aut\'onoma de Madrid, Cantoblanco, 28049 Madrid, Spain
\\[.5ex]
$^2$ Instituto de F\'{\i}sica Corpuscular, Universidad de Valencia and CSIC\\ 
 Edificio Institutos Investigaci\'on, Catedr\'atico Jos\'e Beltr\'an 2, 46980 Spain
\\[.5ex]
$^3$ IJCLab, Pôle Théorie (Bat. 210), CNRS/IN2P3, 91405 Orsay, France 

}

\vspace*{2mm}
\end{center}

\renewcommand*{\thefootnote}{\arabic{footnote}}
\setcounter{footnote}{0}

\begin{abstract}
We analyze how neutrino oscillation and coherent elastic neutrino-nucleus scattering data impact the global SMEFT fit. We first review the mapping between the SMEFT parameters and the so-called NSI framework, commonly considered in the neutrino literature. We also present a detailed discussion of how the measurements for the normalization of neutrino fluxes and cross sections, that will also be affected by the new physics, indirectly impact the measured oscillation probabilities.  
We then analyze two well-motivated simplified scenarios. Firstly, we study a lepton flavour conserving case, usually assumed in global SMEFT analyses, showing the complementarity of neutrino oscillation  and CE$\nu$NS experiments with other low-energy observables. We find that the inclusion of neutrino data allows to constrain previously unbounded SMEFT operators involving the tau flavour and confirm the improvement of the constraint on a combination of Wilson coefficients previously identified. Moreover, we find that neutrino oscillation constraints on NSI are improved when embedded in the global SMEFT framework. Secondly, we study a lepton flavour violating scenario and find that neutrino data also improves over previously derived global constraints thanks to its sensitivity to new combinations of Wilson coefficients.

\end{abstract}

\end{titlepage}

\tableofcontents

\section{Introduction}
\label{sec:Introduction}

In the search for new physics beyond the Standard Model (SM), Effective Field Theories (EFTs) provide a versatile tool to analize data in an agnostic, model-independent way. Through the use of EFTs, these searches can be made with very minimal assumptions and the constraints derived are thus particularly robust. This approach lies at the opposite end of the spectrum and is complementary to searches for a specific new physics model, usually characterized by only a few parameters. This latter strategy will generally provide much stronger constraints, but its applicability will be reduced to the particular model assumed. Conversely, one of the main challenges of the EFT approach is the very large number of free parameters that are, in principle, uncorrelated and should be analyzed globally for the searches and constraints to be truly general and model-independent, maximizing the complementarity with the specific model searches. Indeed, this freedom often allows to move along particular directions in parameter space that avoid the most stringent constraints and significantly relax the global results that may be derived. It is then very interesting to find observables capable of closing these ``flat directions'', which are always the main bottleneck of the truly general and model-independent constraints. 

In this context, the SMEFT~\cite{Buchmuller:1985jz,Grzadkowski:2010es} is the most general EFT that can be built with the SM particle content and respecting its fundamental symmetries. Even stopping the tower of effective operators at dimension 6, 2499 parameters are required to fully characterize the Lagrangian at this level~\cite{Alonso:2013hga}. Thus, the SMEFT programme must analize dozens of different observables simultaneously with the aim of exploring as fully as possible its vast parameter space and derive solid, model-independent new physics constraints. In parallel, flat and poorly constrained directions need to be identified in order to find new observables with the most potential to improve the overall constraints. 
Given the daunting nature of this task, SMEFT analyses are usually performed in subsectors of the theory or with subsets of observables, such as collider measurements, electroweak precision observables or flavour data.

The analogous approach with neutrino oscillation and scattering data is known as Non-Standard neutrino Interactions (NSI). It represents an EFT parametrization of hypothetical new interactions in which neutrinos would be involved, affecting their production, detection and/or propagation, and thus the neutrino oscillation phenomenon~\cite{Wolfenstein:1977ue,Mikheyev:1985zog,Roulet:1991sm,Guzzo:1991hi,Grossman:1995wx,Gonzalez-Garcia:1998ryc,Farzan:2017xzy}. Unlike the SMEFT, the NSI formalism is not an exhaustive list of all EFT operators allowed by the matter content and symmetries under consideration. It is instead a useful parametrization of the effects that would be relevant for the neutrino oscillation phenomenon. As such, it is interesting to study the connection between the SMEFT and NSI formalisms. Several works have already investigated the combinations of SMEFT operators that give rise to the relevant NSI parameters~\cite{Gavela:2008ra,Falkowski:2017pss,Jenkins:2017jig,Altmannshofer:2018xyo,Bischer:2019ttk,Breso-Pla:2023tnz,Cherchiglia:2023aqp}.
One of the main consequences of assuming that neutrino NSI arise from SMEFT operators is that the $SU(2)_L$ gauge invariance built into the SMEFT implies relations between the neutrino NSI operators and processes involving charged leptons, generally much easier to constrain. Thus, the stronger constraints stemming from the charged lepton sector generally lead to the conclusion that NSI originating from heavy mediators\footnote{These constraints may be evaded when NSI are mediated through light mediators, see Refs.~\cite{Wise:2014oea,Farzan:2015doa,Farzan:2015hkd,Farzan:2016wym,Babu:2017olk,Heeck:2018nzc,Proceedings:2019qno,Babu:2019mfe,Farzan:2019xor,Greljo:2021npi,Greljo:2022dwn,Bernal:2022qba} for particular examples.} is too strongly constraint to have an impact in present or near-future neutrino experiments~\cite{Antusch:2008tz,Gavela:2008ra}.

Nevertheless, despite the very strong constraints that other data place on the relevant SMEFT operators for neutrino NSI, some flat or poorly constrained directions exist along which sizable contributions could be possible (see {\it e.g.}~\cite{Falkowski:2017pss,Fernandez-Martinez:2024bxg}). It is therefore interesting to explore the capabilities and complementary of neutrino bounds on NSI to contribute to the global SMEFT constraints. In this work we explore this possibility and find that, in fact, neutrino constraints on NSI from neutrino oscillations and Coherent Elastic neutrino-Nucleus Scattering (CE$\nu$NS)~\cite{Freedman:1973yd,Drukier:1984vhf} do improve upon present constraints from other observables for particular operator combinations or even provide unique constraints on some previously unbounded operators, contributing to the global SMEFT fit.

In Section~\ref{sec:Oscillations} we review how different neutrino experiments are sensitive to neutrino NSI. In this context, we will take particular care to describe how the necessary measurements of neutrino fluxes and cross sections, required to determine the oscillation probability, are also affected by NSI operators and induce an indirect dependence on them. In Section~\ref{sec:Formalism} we present the connection between the NSI and SMEFT formalism and highlight also the connection with operators involving charged leptons. In Sections~\ref{sec:LFC} and~\ref{sec:LFV} we explore how neutrino data may contribute to the global SMEFT constraints in two particular simplified but well-motivated and meaningful scenarios focusing on lepton flavour conserving and violating operators respectively. Finally, in Section~\ref{sec:Conclusions} we conclude and summarize our findings. 

\section{Neutrino oscillation event rates in presence of general NSI}
\label{sec:Oscillations}

In this section we review how neutrino experiments are affected by general NSI, which can alter the experimental observations by modifying either the propagation itself or the production and detection rates~\cite{Wolfenstein:1977ue,Mikheyev:1985zog,Roulet:1991sm,Guzzo:1991hi,Grossman:1995wx,Gonzalez-Garcia:1998ryc,Farzan:2017xzy}. We will pay special attention to how the oscillation rates must be normalized so as to compare them with the actual measurement at experiments, an issue that is sometimes overlooked. 
Finally, we also briefly review CE$\nu$NS in presence of NSI, since it plays a fundamental role breaking the degeneracies from an analysis with only oscillation data. 

We parameterize new physics effects at neutrino oscillation and scattering experiments by means of the most general NSI Lagrangian, which can be classified as neutral current (NC) and charged current (CC) NSI. In the SM flavour basis they read:
\begin{align}
    \mathcal{L}_{\text{NC-NSI}}\supset &
    -2\sqrt{2}G_F \nsi{f,X}{\alpha\beta}(\bar{\nu}_\alpha \gamma_\mu P_L\nu_\beta)(\bar{f}\gamma^{\mu}P_Xf)\,,\label{eq:NC-NSI}\\
    \mathcal{L}_{\text{CC-NSI}}\supset &-2\sqrt{2} G_F \bigg\{
    \varepsilon_{\alpha\beta}^{\mu e X}(\bar{\nu}_\alpha \gamma_\mu P_L\nu_\beta)(\bar{\mu}\gamma^{\mu}P_X e)+ \varepsilon_{\alpha\beta}^{ud L}(\bar{e}_\alpha \gamma_{\mu}P_L\nu_\beta)(\bar{u}\gamma^\mu P_L d)
    \nonumber\\
    +&\varepsilon_{\alpha\beta}^{ud R}(\bar{e}_\alpha \gamma_{\mu}P_L\nu_\beta)(\bar{u}\gamma^\mu P_R d)+\tfrac12\varepsilon_{\alpha\beta}^{ud S}(\bar{e}_\alpha P_L\nu_\beta)(\bar{u} d)\nonumber\\
    +&
    \tfrac12\varepsilon_{\alpha\beta}^{ud P}(\bar{e}_\alpha P_L\nu_\beta)(\bar{u}\gamma_5 d)
    +\varepsilon_{\alpha\beta}^{ud T}(\bar{e}_\alpha\sigma_{\mu\nu} P_L\nu_\beta)(\bar{u} \sigma^{\mu\nu}P_Ld)
    \bigg\}+h.c.
    \label{eq:CC-NSI}
\end{align}
Here, $G_F$ is extracted from $\mu$ decay, the index \( X \) denotes the Lorentz structure of the operator, taking values \( X = L, R \), and $P_{L, R} = \frac{1}{2}(1 \mp \gamma_5)$ are the chirality projection operators\footnote{Notice that for NC NSI Eq.~\eqref{eq:NC-NSI} is the only possible Lorentz structure. Scalar or tensor bilinears would require either the addition of right-handed neutrinos or violating $L$, which would in turn require to go to higher order in the SMEFT matching to restore gauge invariance through Higgs insertions. See Refs.~\cite{Ge:2018uhz,Babu:2019iml,Smirnov:2019cae} for models in which they arise from light mediator exchanges.}.
For convenience, we can also define the corresponding vector and
axial-vector combinations of NC NSI coefficients as:

\begin{equation}
\label{eq:NSIaxialvector}
 \varepsilon_{\alpha \beta}^{f, V} \equiv \varepsilon_{\alpha\beta}^{f,L} +\varepsilon_{\alpha\beta}^{f,R} \;\;{\rm{and}} \;\;  \varepsilon_{\alpha \beta}^{f, A} \equiv \varepsilon_{\alpha\beta}^{f,L} - \varepsilon_{\alpha\beta}^{f,R}\,.
\end{equation}

In general, CC NSI, which typically impact the neutrino production and detection processes, and NC NSI, which are rather mostly relevant for neutrino propagation inducing non-standard matter effects, are expected to appear simultaneously. Indeed, several of the SMEFT operators that may induce NSI do contribute to both the CC and NC operators, as we will explicitly show in the next section. Even beyond the SMEFT approach, it is SU(2)$_L$ gauge invariance what relates CC and NC NSI. As such, in all generality, NSI of both CC and NC types should be considered simultaneously when analyzing data.

Nevertheless, while both CC and NC NSI will impact the neutrino oscillation phenomenon, the impact of CC NSI does not require neutrinos to propagate over long baselines. As such, this ``zero distance effect'' may be exploited to set stringent bounds on these types of NSI~\cite{Biggio:2009kv,Falkowski:2021bkq}.
On the other hand, the same zero distance effect will impact any data used to estimate the fluxes and cross sections needed to extract the oscillation probability from the number of events measured. When this indirect dependence on NSI is properly accounted for, the zero distance effect as well as the dependence on Lepton Flavour Conserving (LFC) CC NSI expected in the oscillation probabilities may cancel in disappearance channels. We elaborate these discussions in the following sections.

\subsection{Charged Current NSI}
\label{sec:CC}

We start discussing how the modification of production and detection rates due to CC NSI affect neutrino oscillation experiments. To begin with, a very important remark is that the neutrino oscillation probability is not a physical observable and has to be carefully defined in presence of new physics effects that impact the production and/or detection processes. Indeed, the standard definition of the oscillation probability $P_{\alpha \beta} (L)$ and its connection to the observable number of events is usually given by:
\begin{equation}
  R_{\alpha}^{{\rm CC} , \beta}(L) = N_T \Phi_\alpha P_{\alpha \beta} (L) \frac{d\sigma_\beta}{d E_\nu}~,
\label{eq:normalprob}
\end{equation}
where $R_{\alpha}^{{\rm CC} , \beta}(L)$ is the differential event rate (as a function of neutrino energy, $E_\nu$) when an initial flux of neutrinos $\Phi_\alpha$ is detected at a distance $L$ from the neutrino source via CC neutrino interactions with a target containing $N_T$ particles and producing an outgoing charged lepton of flavour $\beta$. 
Here, the initial flux $\Phi_\alpha$ is produced in a decay of a parent particle in association with a charged lepton of flavour $\alpha$, while the differential cross section of the detection process as $\nu T \rightarrow \ell_\beta + \ldots$ would be given by $\frac{d\sigma_\beta}{d E_\nu}$.
Thus, the neutrino oscillation probability is extracted from the measured number of events at a given experiment as:
\begin{equation}
P_{\alpha \beta} (L) = \frac{R_{\alpha}^{{\rm CC} , \beta}(L)}{{N_T \Phi_\alpha^{\mathrm{est}} \frac{d\sigma_\beta^{\mathrm{est}}}{d E_\nu}}}\,,
\label{eq:generalprob}
\end{equation}
where $\Phi_\alpha^{\mathrm{est}}$ and $\sigma_\beta^{\mathrm{est}}$ are the estimations of the neutrino flux and cross sections obtained by the corresponding experimental collaboration.
As discussed below, the way these quantities are estimated impacts the actual sensitivity to new physics parameters. 

Neutrino oscillations take place and $P_{\alpha \beta} (L) \neq \delta_{\alpha \beta}$ because the neutrino flavour eigenstates which are produced and detected do not coincide with the propagation eigenstates and are related through the mixing matrix $U_{\alpha i}$, leading to the well-known neutrino oscillation probability:
\begin{equation}
P_{\alpha \beta}(L)  = \sum_{i,j}  U_{\alpha i}^* U_{\beta i}^{} U_{\alpha j}^{} U_{\beta j}^*\, e^{-i\frac{\Delta m^2_{ij} L}{2E}}\,.
\label{eq:usualprob}
\end{equation}
The matrix elements $U_{\alpha i}^*$ and $U_{\beta i}^{}$ respectively stem from the initial flux $\Phi_\alpha$ and detection cross section $\sigma_\beta$ amplitudes when the neutrino fields are expressed in the propagation eigenbasis. Nevertheless, since $\sum_i |U_{\alpha i}|^2 = 1$, the factorization performed in Eq.~\eqref{eq:normalprob} is possible and $\Phi_\alpha$ and $\sigma_\beta$ are unaffected even if the elements $U_{\alpha i}$ and $U_{\beta i}^*$ are ascribed to the oscillation probability $P_{\alpha \beta}(L) $ instead. As such, the oscillation probability can be cleanly extracted from Eq.~\eqref{eq:generalprob}. 

This is no longer the case when new physics effects impact the production and/or detection processes. 
For illustration purposes, we first focus on a concrete simplified example of NSI affecting exclusively neutrino production through pion decay. 
In presence of CC NSI affecting $\pi \to \mu \nu_j$, the amplitude would be given by:
\begin{equation}
    \mathcal{M} (\pi^+ \to \mu^+ \nu_j)  
= 
 - i \, m_\mu   f_{\pi}  \frac{V_{ud}}{v^2} \sum_\alpha \left(\delta_{\mu \alpha} + \varepsilon^\pi_{\mu \alpha} \right) U_{\alpha j}^* (\bar{u}_{\nu_j} P_L v_{\mu})\,,
 \label{eq:piodecay}
\end{equation}
where $V_{ud}$ stands for the corresponding CKM matrix element and  $f_\pi$ is the pion decay constant. Here, the $\delta_{\mu\alpha}$ accounts for the SM contribution to this process (which arises from an axial operator), while
\begin{equation}
\varepsilon^\pi_{\mu \alpha} \equiv \varepsilon_{\mu \alpha}^{ud A} +\frac{m_\pi^2}{m_\mu\left(m_u+m_d\right)} \varepsilon_{\mu \alpha}^{ud P}\,, 
\end{equation}
may include new contributions from axial operators as well as a pseudoscalar contribution (in fact, notice the significant enhancement of the pseudoscalar contribution with the light fermion masses $m_\mu$, $m_u$ and $m_d$ with respect to the axial contributions, where a chirality flip is needed). 

Importantly, from Eq.~\eqref{eq:piodecay} we see that now $\sum_\alpha (\delta_{\mu \alpha} + \varepsilon^\pi_{\mu \alpha} ) U_{\alpha i}^*$ replaces $U_{\alpha i}^*$ in the standard scenario. However, since $\sum_{\alpha} | \delta_{\mu \alpha} + \varepsilon^\pi_{\mu \alpha} |^2 \neq 1$, not only the combination of propagation eigenstates produced at the source but also the neutrino flux itself $\Phi_\alpha$ is affected. As such, special care must be taken in considering not only the non-trivial dependence that $\Phi_\alpha$ (and $\sigma_\beta$ when detection is also affected) may have on the new physics, but also whether the estimations $\Phi_\alpha^{\mathrm{est}}$ and $\sigma_\beta^{\mathrm{est}}$ are affected too, and propagate this dependence accordingly in Eq.~\eqref{eq:generalprob}. These will greatly depend on the concrete experimental setup under consideration.

The most common scenario when considering neutrinos from pion decay corresponds to pion beams decaying in a decay pipe towards the neutrino detector. The neutrino flux $\Phi_\alpha$ will in this case depend on the pion flux $\Phi_\pi$ itself. The pion flux is often estimated either through Monte Carlo methods or through direct measurements of hadroproduction. In any event, pion production from the proton beam impinging the target will not be affected by the NSI under consideration here. Moreover, if the decay pipe is long enough, all pions will decay. As such, the neutrino flux from pion decay associated with a charged lepton of a flavour $\alpha=\mu$ will be:
\begin{equation}
    \Phi_\alpha = \Phi_\pi \mathrm{Br}(\pi \to \mu \nu).
\end{equation}

Notice that in general the charged lepton flavour would not be detected, so neutrinos from $\pi \to e \nu$ could also contribute if enhanced by NSI. 
Nevertheless, measurements of this branching ratio are in good agreement with the SM expectations and the corresponding lepton flavour universality constraint from the ratio of both decay modes may be used to derive strong constraints on these NSI instead~\cite{Biggio:2009nt,Falkowski:2021bkq}. As such, $\pi \to e \nu$ can be neglected for simplicity and it is a good approximation to assume that $\pi \to \mu \nu$ has a $100\%$ decay rate for our purposes. In any case, our conclusions are not affected by this assumption. 
In this context, to compensate the $\sum_\alpha \left(\delta_{\mu \alpha} + \varepsilon^\pi_{\mu \alpha} \right) U_{\alpha i}^*$ replacing $U_{\alpha i}^*$ in the produced eigenstate, we need to normalize with $\sum_\alpha \left| \delta_{\mu \alpha}  + \varepsilon^\pi_{\mu \alpha} \right|^2$ so that $\sum_\alpha P_{\mu \alpha}(L=0)=1$, that is, for each pion decay a neutrino is produced. Thus the oscillation probability reads:
\begin{equation}
P_{\mu \gamma}(L)  =\frac{\sum_{\alpha,\beta,i,j} \left(\delta_{\mu \alpha}  + \varepsilon^\pi_{\mu \alpha} \right) U_{\alpha i}^* U_{\gamma i}^{} \left(\delta_{\mu \beta}  + \varepsilon^{\pi*}_{\mu \beta} \right) U_{\beta j}^{} U_{\gamma j}^*\, e^{-i\frac{\Delta m^2_{ij} L}{2E}}}{\sum_\alpha \left| \delta_{\mu \alpha}  + \varepsilon^\pi_{\mu \alpha} \right|^2}\,.
\label{eq:correctprob}
\end{equation}
Interestingly, this oscillation probability already displays observable effects at vanishing baseline $L$. This ``zero distance'' effect will be given by:
\begin{equation}
P_{\mu \gamma}(L=0) 
 =\frac{ \left|\delta_{\mu \gamma}  + \varepsilon^\pi_{\mu \gamma} \right|^2}{\sum_\alpha \left| \delta_{\mu \alpha}  + \varepsilon^\pi_{\mu \alpha} \right|^2}~.
\label{eq:correctzerodist}
\end{equation}
In disappearance channels, the leading order reduces to:
\begin{equation}
P_{\mu \mu}(L=0) \simeq 1-|\varepsilon^\pi_{\mu e}|^2-|\varepsilon^\pi_{\mu \tau}|^2\,,
\label{eq:correctzerodistdisap}
\end{equation}
and the would-be leading linear contribution with $\varepsilon^\pi_{\mu \mu}$ is lost since we have properly accounted for the normalization in the branching ratio through the denominator of Eq.~\eqref{eq:correctprob}. This subtlety has however sometimes been missed in the literature and misused to set strong constraints on diagonal CC NSI.
Only a subleading (quadratic) dependence on the off-diagonal elements $\varepsilon^\pi_{\mu \alpha}$ remains, nevertheless these are more efficiently probed by appearance channels:
\begin{equation}
P_{\mu \alpha}(L=0) \simeq  |\varepsilon^\pi_{\mu \alpha}|^2\,, \quad {\rm for}~ \alpha\neq\mu\,.
\label{eq:correctzerodistap}
\end{equation}

Thus, the zero distance effect at appearance channels provides a very effective way of constraining CC NSI and their impact at production and detection. In fact, we will make use of this effect in Section~\ref{sec:LFV} to improve upon present constraints on Lepton Flavour Violating (LFV) SMEFT operators~\cite{Fernandez-Martinez:2024bxg} using bounds on $\nu_\mu \to \nu_e$ transitions at zero distance.

Conversely, Lepton Flavour Conserving (LFC) CC NSI in general cannot be probed by searching for zero distance effects in disappearance channels. 
This is particularly relevant for the flavour conserving SMEFT scenario, as often adopted in many analyses. Indeed, since only diagonal NSI $\varepsilon_{\alpha \alpha}$ would be present, all sensitivity to production and detection CC NSI (even beyond the zero distance effect) is lost~\cite{Breso-Pla:2023tnz,Cherchiglia:2023aqp} in Eq.~\eqref{eq:correctprob}.
This implies that, while in general it is not consistent to analyze neutrino oscillations in presence of only propagation NC NSI, it may be justified when studying a framework with only LFC NSI. We will make use of this observation in Section~\ref{sec:LFC} where we will combine a global fit of LFC SMEFT operators~\cite{Breso-Pla:2023tnz} with the constraints on LFC NC NSI from a global fit including only NC and not CC NSI~\cite{Coloma:2023ixt}. 

The lack of sensitivity to LFC CC NSI stemming from a proper normalization of the probabilities when in presence of CC NSI is more generic\footnote{Also in presence of a non-unitary mixing matrix whose impact in oscillations can be mapped to the NSI formalism, see Refs.~\cite{Blennow:2016jkn,Aloni:2022ebm}.} and goes beyond the simplified example we studied so far. Indeed, even if we consider a setup in which the decay pipe is not long enough for all pions to decay and thus the neutrino flux is now proportional to the width instead of to the branching ratio, a measurement of the pion decay width would now be needed for the estimation of the neutrino flux $\Phi_\alpha^{\mathrm{est}}$. This measurement will be proportional to the modulus squared of the amplitude in Eq.~\eqref{eq:piodecay}, that is:
\begin{equation}
    \Phi_\alpha^{\mathrm{est}} = \Phi_\alpha \sum_\alpha | \delta_{\mu \alpha}  + \varepsilon^\pi_{\mu \alpha} |^2~.
    \label{eq:fluxnorm}
\end{equation}
Thus, even if the normalization from the branching ratio is not present, upon substituting Eq.~\eqref{eq:fluxnorm} in Eq.~\eqref{eq:generalprob}, Eq.~\eqref{eq:correctprob} is again obtained. 

In more general terms beyond our pion decay example, the differential event rate in Eq.~\eqref{eq:normalprob} will generally be given by:
\begin{equation}
R_{\alpha}^{{\rm CC}, \beta}(L) = \frac{N_SN_T}{32 \pi L^2 E_S m_T E_\nu} \sum_{k, l} e^{-i \frac{\Delta m_{k l}^2 L}{2 E_\nu}} \int d \Pi_{P} \mathcal{M}_{\alpha k}^P \overline{\mathcal{M}}_{\alpha l}^P \int d \Pi_D \mathcal{M}_{\beta k}^D \overline{\mathcal{M}}_{\beta l}^D\,,
\label{eq:Ralpha}
\end{equation}
where $N_{S, T}$ are the number of source and target particles, $E_{S}$ the energy of the monochromatic flux of the source particle and $m_{T}$ the mass of the target particle (which is considered to be at rest). Note that the $E_\nu$ is uniquely determined by $E_S$ in the case of a two-body decay, which we are assuming for simplicity\footnote{We point the interested reader to Ref.~\cite{Falkowski:2021bkq} for a more general scenario.}. Now, in the presence of the CC NSI defined in Eq.~\eqref{eq:CC-NSI}, the production and detection amplitudes at linear order in the NSI parameters are given by:
\begin{equation}
\begin{aligned}
& \mathcal{M}_{\alpha k}^P=U_{\alpha k}^* A_L^P+\sum_X\left[\varepsilon^X U\right]_{\alpha k}^* A_X^P\,, \\
& \mathcal{M}_{\beta k}^D=U_{\beta k} A_L^D+\sum_X\left[\varepsilon^X U\right]_{\beta k} A_X^D\,, 
\end{aligned} 
\label{eq:Mdet_betak}
\end{equation}
where $A^P_X$ and $A^D_X$ are the relevant amplitudes for the Lorentz structure $X\equiv L, R, S, T, P$ of the corresponding operator contributing to the production or detection process respectively. 

As discussed above, the estimations of $\Phi_\alpha^{\mathrm{est}}$ and $\sigma_\beta^{\mathrm{est}}$ are very facility-dependent and will necessarily be based on some measurement. As such, they also consequently depend on the new physics parameters and are \emph{not} ``pure'' SM predictions\footnote{Even though they are defined as such in some references.}. This is one of the crucial points where new physics dependence enters the normalization of the oscillation probability, potentially cancelling the corresponding sensitivity. 
Let us define two broad scenarios depending on how $\Phi_\alpha^{\mathrm{est}}$ and $\frac{d\sigma_\beta^{\mathrm{est}}}{d E_\nu}$ are estimated:

\begin{itemize}
\item {\bf Underlying SM structure assumed.} It is common when estimating fluxes and cross sections through simulations to assume the SM weak interactions to generate events. This is also a usual assumption in phenomenological works (see for instance Refs.~\cite{Breso-Pla:2023tnz,Cherchiglia:2023aqp}). In this scenario, the estimated flux and cross section would then be: 
\begin{equation}
\Phi_\alpha^{\mathrm{est}} = N_\Phi(\varepsilon) \frac{N_S \int d \Pi_{P^{\prime}}\left|A_L^P\right|^2}{8 E_S \pi L^2}, \quad \frac{d\sigma_\beta^{\mathrm{est}}}{dE_\nu} = N_\sigma(\varepsilon) \frac{\int d \Pi_D\left|A_L^D\right|^2}{4 E_\nu m_T} \, .
\label{eq_est1}
\end{equation}
Here $N_\Phi(\varepsilon)$ and $N_\sigma(\varepsilon)$ take into account that, even though the SM operators are usually assumed in Monte Carlo simulations, their normalization (branching ratios of corresponding processes, relevant form factors for scattering amplitudes, etc) will come from measurements and, as such, will be affected by the new physics parameters $\varepsilon$ too\footnote{It can be tempting to neglect this $\varepsilon$ dependence against the, usually sizable, systematic uncertainties affecting these normalizations. However, this $\varepsilon$ dependence is of the \emph{same} order as the one considered in Eq.~\eqref{eq:Mdet_betak}. As such, it is not consistent to neglect one and not the other, and doing so may lead to wrong expectations for the sensitivity to  $\varepsilon$.}. Upon substituting Eq.~\eqref{eq_est1} in Eq.~\eqref{eq:generalprob} we obtain:
\begin{equation}\label{eq:prob_est1}
\begin{aligned}
P_{\alpha \beta} (L) &= \frac{\sum_{k, l} e^{-i \frac{\Delta m_{k l}^2 L}{2 E_\nu}}}{N_\Phi(\varepsilon)N_\sigma(\varepsilon)}   \\
& \times\left[U_{\alpha k}^* U_{\alpha l}+p_{X L}\left(\varepsilon^X U\right)_{\alpha k}^* U_{\alpha l}+p_{X L}^* U_{\alpha k}^*\left(\varepsilon^X U\right)_{\alpha l}+p_{X Y}\left(\varepsilon^X U\right)_{\alpha k}^*\left(\varepsilon^Y U\right)_{\alpha l}\right] \\
& \times\left[U_{\beta k} U_{\beta l}^*+d_{X^{\prime} L}\left(\varepsilon^{X^{\prime}} U\right)_{\beta k} U_{\beta l}^*+d_{X^{\prime} L}^* U_{\beta k}\left(\varepsilon^{X^{\prime}} U\right)_{\beta l}^*+d_{X^{\prime} Y^{\prime}}\left(\varepsilon^{X^{\prime}} U\right)_{\beta k}\left(\varepsilon^{Y^{\prime}} U\right)_{\beta l}^*\right],
\end{aligned}
\end{equation}
where in this formula repeated $X, Y, X^{\prime}, Y^{\prime}$ indices are implicitly summed over, and we define the production and detection coefficients as
\begin{equation}
p_{X Y} \equiv \frac{\int d \Pi_{P} A_X^P \bar{A}_Y^P}{\int d \Pi_{P}\left|A_L^P\right|^2}\,, 
\quad 
d_{X Y} \equiv \frac{\int d \Pi_D A_X^D \bar{A}_Y^D}{\int d \Pi_D\left|A_L^D\right|^2}\, .
\end{equation}
If the data used to normalize the event rates in the simulations comes from the exact same processes through which neutrinos are respectively produced and detected in Eq.~\eqref{eq:Ralpha}, then:
\begin{equation}
\begin{aligned}
N_\Phi(\varepsilon)&=1+p_{X L}\left(\varepsilon^{X}\right)_{\alpha \alpha}^* +p_{X L}^* \left(\varepsilon^{X}\right)_{\alpha \alpha}+p_{X Y}\left(\varepsilon^{X*}  \varepsilon^Y\right)_{\alpha \alpha}, \\
 N_\sigma(\varepsilon)&= 1+d_{X^{\prime} L}\left(\varepsilon^{X^{\prime}}\right)_{\beta \beta} +d_{X^{\prime} L}^* \left(\varepsilon^{X^{\prime}}\right)_{\beta \beta}^*+d_{X^{\prime} Y^{\prime}}\left(\varepsilon^{X^{\prime}}  \varepsilon^{Y^{\prime}*}\right)_{\beta \beta}.
\end{aligned} 
\label{eq:normnorm}
\end{equation}
Thus, when the neutrino flux and cross section are normalized through measurements of the same processes involved in $R_{\alpha}^{{\rm CC}, \beta}$, then $P_{\alpha \alpha} (0) = 1$ and the zero distance in disappearance channels is lost at linear order in $\varepsilon$.

In general, the situation will be more complex than the simplified generalized scenario we have presented here. 
Beyond our approximation from a two-body decay, $p_{X Y}$ and $d_{X Y}$, and hence $N_\Phi(\varepsilon)$ and $N_\sigma(\varepsilon)$, may have a non-trivial energy dependence\footnote{The energy dependence of $N_\Phi(\varepsilon)$ and $N_\sigma(\varepsilon)$ will depend on the data used to normalize the Monte Carlo predictions and it may or not be the same as for the actual measurement. Thus, the level of cancellation will strongly depend on this and in some instances energy resolution may be exploited to derive constraints on NSI through disappearance channels (see~\cite{Falkowski:2019xoe}).}.
Moreover, many inputs go into the computation $\Phi_\alpha^{\mathrm{est}}$ and $d\sigma_\beta^{\mathrm{est}}$ generally involving clever combinations of Monte Carlo techniques (usually assuming underlying SM interactions) but calibrated against data (that would then be affected by NSI). Thus, in practice special care must always be taken to correctly propagate this NSI dependence into $\Phi_\alpha^{\mathrm{est}}$ and $d\sigma_\beta^{\mathrm{est}}$ and hence to the oscillation probability extracted from the observed number of events through Eq.~\eqref{eq:generalprob}.

\item {\bf Normalization with a near detector.}

Even though the former scenario is more often assumed, most modern neutrino oscillation experiments actually calibrate the expectations for the unoscillated number of events through measurements at a near detector before the onset of oscillations. The goal is that the measured events at the near detector mimic as much as possible the production and detection processes at the far detector so as to cancel systematic uncertainties to the largest possible extent. As such, some of the production and detection effects caused by NSI will necessarily cancel together with the systematics. However, since oscillations have not developed yet at the near detector, the corresponding measurement is normally done through the original flavour also when an appearance channel is searched for at the far detector. As such, in this scenario the normalization factors for the oscillation probability in Eq.~\eqref{eq:prob_est1} will be given by:
\begin{equation}
\begin{aligned}
N_\Phi(\varepsilon)&=1+p_{X L}\left(\varepsilon^{X}\right)_{\alpha \alpha}^* +p_{X L}^* \left(\varepsilon^{X}\right)_{\alpha \alpha}+p_{X Y}\left(\varepsilon^{X*}  \varepsilon^Y\right)_{\alpha \alpha}, \\
 N_\sigma(\varepsilon)&= 1+d_{X^{\prime} L}\left(\varepsilon^{X^{\prime}}\right)_{\alpha \alpha} +d_{X^{\prime} L}^* \left(\varepsilon^{X^{\prime}}\right)_{\alpha \alpha}^*+d_{X^{\prime} Y^{\prime}}\left(\varepsilon^{X^{\prime}}  \varepsilon^{Y^{\prime}*}\right)_{\alpha \alpha}.
\end{aligned} 
\label{eq:normnorm2}
\end{equation}
Thus, in this scenario the zero distance effect is not observable in disappearance channels either.

As the previous scenario, this description is also somewhat oversimplified. Indeed, even if the flux and cross section measured in the near detector try to mimic as much as possible those in the far to cancel systematic uncertainties, this is never exactly the case. For instance, while the far detector sees a point source, at the near detector the neutrinos come from an extended source with different off-axis angles and a necessarily different spectrum. Similarly, as mentioned above, the neutrino flavour detected may be different in the near and far detectors, and the cross sections may vary accordingly. In order to correctly estimate the events at the far detector from the near detector measurements, these effects need to be corrected for and when these corrections are implemented SM physics is assumed to relate both measurements. As such, there will be some distortion to the simple expressions of Eq.~\eqref{eq:normnorm2}, preventing a complete cancellation of the effects.  These corrections would need to be quantified for each facility under study and included if relevant.

\end{itemize}

In summary, the general observation is that, when the indirect dependence on NSI parameters from the modification of the neutrino fluxes and cross sections is properly accounted for, the sensitivity to exclusively LFC CC NSI (that is, diagonal $\varepsilon_{\alpha \alpha}$) is lost (see Eq.~\eqref{eq:correctprob}) together with the leading contribution of the zero distance effect in disappearance channels. In order to avoid this cancellation, an estimation for the fluxes $\Phi_\alpha^{\mathrm{est}}$ and cross sections $\sigma_\beta^{\mathrm{est}}$ unaffected by the NSI would be needed. However, these estimations necessarily rely on some measurements that will generically be affected by the NSI (see Eq.~\eqref{eq:fluxnorm}), in a similar way to the oscillation process, hence partially cancelling its dependence. 

The estimation of the neutrino cross sections is often more subtle, as it does not always entail a measurement of the same process taking place at the neutrino detector and is often estimated from Monte Carlo simulations. Nevertheless, the data used to calibrate these simulations will generally also be affected by NSI. For instance, one of the most basic inputs to estimate these CC interactions will be the CKM matrix element $V_{ud}$, which is extracted from superallowed $\beta$ decays that are affected by the same CC NSI defined in Eq.~\eqref{eq:CC-NSI} and affects neutrino CC interactions in the detection process. Similar arguments apply to the extraction of the different relevant form factors. While the combinations in which the NSI appear in each of these processes may vary, avoiding an exact cancellation since the normalization factors would not be given by Eq.~\eqref{eq:normnorm} in this case, we will argue in Section~\ref{sec:LFC} that the NSI involved are, in any event, either more strongly constrained through direct measurements than the current sensitivity of oscillation experiments or they only appear at a subleading (quadratic) order and their impact can be neglected. Thus, it is meaningful to neglect the impact of CC NSI in neutrino oscillations when studying a LFC scenario. Therefore, in the following we will focus on discussing what combinations of NC NSI operators neutrino experiments are sensitive to.

\subsection{Neutral Current NSI}

The presence of NC NSI may affect both neutrino propagation in matter (if the new operators are vector-like) and the detection cross section via NC interactions in the detector (for any Lorentz structure considered). Here we consider the two effects separately just to simplify the discussion; however we note that in general the two may be present for a given experiment simultaneously.

\subsubsection{Impact of NC NSI in detection processes}\label{subsec:NC-NSI-Det}

In case of neutral-current interactions, the event rates should account for the interactions of all neutrinos that arrive to the detector, regardless of their flavour. In fact, since there is no outgoing charged lepton in the detection process, in this case it is more convenient to compute the differential event rates in the neutrino mass basis\footnote{This also circumvents the problem of a potentially ill-defined flavour basis when also in presence of CC NSI.}. While the new physics effects in the interaction vertex may still be expressed in the flavour basis (as is commonly done in the literature for neutrino NSI), it is possible to write down new amplitudes for the detection process in the mass basis, in full analogy to Eq.~\eqref{eq:Mdet_betak}, as:  
\begin{equation}
\mathcal{M}_{j k}^D \equiv \sum_\beta U^*_{\beta j} \mathcal{M}_{\beta k}^D = \delta_{jk} A_L^D + 
\sum_X \left[U^\dagger \varepsilon^X U \right]_{j k} A_X^D~ \, .
\label{eq:Mdet_jk}
\end{equation}
For oscillations in vacuum, the differential event rates with an outgoing neutrino $\nu_j$ therefore read 
\begin{equation} 
R_{\alpha}^{{\rm NC}, j} =
\frac{N_T}{32 \pi L^2 m_S m_T E_\nu} \sum_{k, l} e^{-i \frac{\Delta m_{k l}^2 L}{2 E_\nu}} \int d \Pi_{P} \mathcal{M}_{\alpha k}^P \overline{\mathcal{M}}_{\alpha l}^P \int d \Pi_D \mathcal{M}_{j k}^D \overline{\mathcal{M}}_{j l}^D \, ,
\end{equation}
while the observed event rates are obtained as the sum of all contributions for the outgoing neutrino mass eigenstates:
\begin{equation}
    R_{\alpha}^{{\rm NC}}  = \sum_j R_{\alpha}^{{\rm NC}, j} \, .
\end{equation}

In the oscillation experiments we consider on the following sections, neutral-current interactions are only relevant for SNO~\cite{SNO:2005oxr,SNO:2006odc,SNO:2008gqy,SNO:2011hxd}, as these induce the breakup of deuterium\footnote{There are also strong constraints on sterile neutrinos~\cite{MINOS:2020llm} from NC measurements in MINOS and MINOS+ which could be used to constrain NSI but are not considered here.}. As explained in detail in  Refs.~\cite{Bahcall:1988em,Bernabeu:1991sd,Chen:2002pv}, this process is mediated by a Gamow-Teller operator. Therefore,  the number of events is sensitive to the axial isovector hadronic current, which in the SM is given by $g_A \equiv g_A^u - g_A^d$. Assuming that the nuclear corrections to $g_A$ are the same in presence of NSI (so they can be factorized out), this implies that the SNO data is sensitive to the same combination of axial NSI operators~\cite{Davidson:2003ha}: $\varepsilon^{u,A}_{\alpha\beta} - \varepsilon^{d,A}_{\alpha\beta}$. Here we will use the bound derived from the numerical analysis of SNO data performed in Ref.~\cite{Coloma:2023ixt}.

Finally, neutral-current interactions are also relevant for \cevns~experiments, which were also included in the global fit to neutrino data in Ref.~\cite{Coloma:2023ixt}. In this case, the impact of NSI operators can be parametrized through a generalized weak charge of the nucleus:
\begin{equation}
  \label{eq:Qweak}
  \mathcal{Q}_{\alpha\beta}
  = Z \big(g_p^V \delta_{\alpha\beta} + \varepsilon_{\alpha\beta}^{p,V} \big)
  + N \big(g_n^V \delta_{\alpha\beta} + \varepsilon_{\alpha\beta}^{n,V} \big),
\end{equation}
with $Z$ and $N$ the number of protons and neutrons in the nucleus upon which neutrinos scatter. $g_p^V = 1/2 - 2\sin^2\theta_w$ and  $g_n^V = -1/2$ are the SM contribution while $\varepsilon _{\alpha\beta}^{p,V} \equiv 2\varepsilon _{\alpha\beta}^{u,V} +
\varepsilon _{\alpha\beta}^{d,V}$ and $\varepsilon _{\alpha\beta}^{n,V} \equiv
2\varepsilon _{\alpha\beta}^{d,V} + \varepsilon _{\alpha\beta}^{u,V}$ are the NSI contributions to neutrino vector couplings to protons and neutrons. The CE$\nu$NS scattering cross section of an incoming $\nu_\alpha$ would thus be given by:
\begin{equation}
  \label{eq:xsec-SM}
  \frac{d \sigma_{\alpha }^\text{coh} (E_R, E_\nu)}{d E_R}
  = \frac{G_F^2}{2\pi} \,
 \sum_\beta |\mathcal{Q_{\alpha \beta}}|^2 \, F^2(q^2) \, m_A
  \bigg(2 - \frac{m_A E_R}{E_\nu^2} \bigg),
\end{equation}
where $E_R$ is the nucleus recoil energy, $m_A$ is the mass of the nucleus and $F(q^2)$ is its nuclear
form factor evaluated at the squared momentum transfer of the process,
$q^2 = 2 m_A E_R$. 

\subsubsection{Impact of NC NSI on the matter potential in neutrino oscillations}
\label{sec:NSIprop}

Regarding the impact of NC NSI in neutrino propagation and the sensitivity of neutrino oscillation experiments we follow Ref.~\cite{Coloma:2023ixt}. In particular, the combination of NSI parameters that determines the matter effects in Earth's interior, with the corresponding weights from its relative amount of electrons, protons and neutrinos, is given by:
\begin{equation}
    \nsi{\oplus}{\alpha\beta}=\nsi{e,V}{\alpha\beta}+(2+Y_n^{\oplus})\nsi{u,V}{\alpha\beta} + (1+2Y_n^{\oplus})\nsi{d,V}{\alpha\beta}\,,
    \label{eq:effective_Earth_NSI}
\end{equation}
where $Y_n^{\oplus}$ is the neutron/proton ratio and can be taken to be constant to very good approximation.
The PREM model~\cite{Dziewonski:1981xy} fixes $Y_n^{\oplus}= 1.012$ in the Mantle and $Y_n^{\oplus}= 1.137$ in the Core, with an average value $Y_n^{\oplus}= 1.051$ over the whole Earth. 

Notice that, since neutrino oscillations are only sensitive to relative phase differences (and hence the sensitivity to mass squared differences), oscillation experiments affected by propagation NSI through the matter potential are only sensitive to the differences of the LFC NSI, namely $\nsi{\oplus}{ee}-\nsi{\oplus}{\mu\mu}$ and $\nsi{\oplus}{\tau\tau}-\nsi{\oplus}{\mu\mu}$.

As previously stated, we will combine the results of the global fit to NSI from neutrino oscillation and CE$\nu$NS data presented in Ref.~\cite{Coloma:2023ixt}. In particular, we will make use of the results of the parameter scan projected to the $\nsi{\oplus}{\alpha\beta}$ space. Notice that CE$\nu$NS and some oscillation experiments such as those observing solar neutrinos are sensitive to different combinations of NSI. Nevertheless, since part of the sensitivity is provided through matter effects through the Earth, it is convenient to project to this subspace the global NSI bounds.

\section{NSI from SMEFT}
\label{sec:Formalism}

The SMEFT is the most general EFT framework built to study new physics heavier than the electroweak scale. Thus, it is introduced as a tower of higher dimensional operators built from the SM particle content and respecting its symmetries:
\begin{equation}\label{eq:LSMEFT}
    \mathcal L_{\rm SMEFT} =  \mathcal{L}_{\rm SM} +\mathcal{L}_{d=5} +  \mathcal{L}_{d=6}+...\,,
\end{equation}
where the $d=N$ operators are suppressed by powers $1/\Lambda^{N-4}$ of the new physics scale $\Lambda$. Therefore, beyond the Weinberg $d=5$ operator that leads to neutrino mass generation, the naively least suppressed new physics effects will be driven by the $d=6$ contributions, which for convenience we normalize as
\begin{equation}
    \mathcal{L}_{d=6} = \sum_i \frac{c_{i} }{v^2}\mathcal{O}_i\,,
\end{equation}
where $v^2=1/\sqrt{2}G_F$, $G_F$ is extracted from $\mu$ decay and $c_i$ are the corresponding dimensionless Wilson Coefficients (WC). We will thus focus our analysis on the impact of the $d=6$ effective operators. In the remainder of the paper we will follow the notation for their coefficients given in~\cite{Breso-Pla:2023tnz}. 

When particularized to the neutrino sector, the SMEFT operators generate at low energies the NSI introduced in the previous section, as has been studied before in the literature, see for instance~\cite{Gavela:2008ra,Falkowski:2017pss,Jenkins:2017jig,Bischer:2019ttk,Breso-Pla:2023tnz,Cherchiglia:2023aqp}.
In this section we review these matching conditions relevant for our study. 

In general, the NSI defined in Eqs.~\eqref{eq:NC-NSI} and~\eqref{eq:CC-NSI} could receive two types of contributions. On the one hand, they can be directly matched to the corresponding 4-fermion SMEFT operator with the same Lorentz structure and flavour indices. On the other hand, they could be generated after integrating out the $W$ and $Z$ bosons, inheriting a dependence on the SMEFT operators that modify the coupling of these bosons to fermions. Here we show both these contributions to NC and CC NSI. We have, however, particularized their corresponding expressions to the cases of flavour conserving NC NSI and flavour violating CC NSI for simplicity. Indeed, as we will argue below, these are the scenarios where we find it easier and more relevant to study the interplay between neutrino oscillation constraints on NSI and the SMEFT program, and will thus be the main focus of Sections~\ref{sec:LFC} and~\ref{sec:LFV} respectively. With this particularization, the matching between dimension-6 SMEFT operators to the NC NSI in Eq.~\eqref{eq:NC-NSI} reads~\cite{Falkowski:2017pss}:
\begin{align}
    \nsi{e,V}{\alpha\alpha}&=  \delta_{e\alpha}\left(\delta g_L^{We}- \delta g_L^{W\mu} + \frac{1}{2}[c_{\ell\ell}]_{e\mu\mu e}\right)
    -(1-4s_{\rm w}^2)\delta g_L^{Z\nu_\alpha}+\delta g_L^{Ze}+\delta g_R^{Ze} 
    \nonumber\\
        &-\frac{1}{2}\Big([c_{\ell\ell}]_{ee\alpha\alpha}+[c_{\ell e}]_{\alpha\alpha ee}\Big)\,,
        \label{eq:vector_nsi_matching_eV}\\
    \varepsilon_{\alpha\alpha}^{u,V}&=\delta g_L^{Zu} +\delta g_R^{Zu}+\left(1-\dfrac{8}{3}\sw^2\right)\delta g^{Z\nu_\alpha}-\dfrac{1}{2}\left([c_{\ell q}^{(1)}]_{\alpha\alpha11}+[c^{(3)}_{\ell q}]_{\alpha\alpha11}+[c_{\ell u}]_{\alpha\alpha11}\right)\,,
    \label{eq:vector_nsi_matching_uV}\\
    \varepsilon_{\alpha\alpha}^{d,V}&=\delta g_L^{Zd} +\delta g_R^{Zd}-\left(1-\dfrac{4}{3}\sw^2\right)\delta g^{Z\nu_\alpha}-\dfrac{1}{2}\left([c_{\ell q}^{(1)}]_{\alpha\alpha11}-[c^{(3)}_{\ell q}]_{\alpha\alpha11}+[c_{\ell d}]_{\alpha\alpha11}\right)\,,
    \label{eq:vector_nsi_matching_dV}\\
    \varepsilon_{\alpha\alpha}^{u,A}&=\delta g_L^{Zu} -\delta g_R^{Zu}+\delta g^{Z\nu_\alpha}-\dfrac{1}{2}\left([c_{\ell q}^{(1)}]_{\alpha\alpha11}+[c^{(3)}_{\ell q}]_{\alpha\alpha11}-[c_{\ell u}]_{\alpha\alpha11}\right)\,,
    \label{eq:vector_nsi_matching_uA}\\
    \varepsilon_{\alpha\alpha}^{d,A}&=\delta g_L^{Zd} -\delta g_R^{Zd}-\delta g^{Z\nu_\alpha}-\dfrac{1}{2}\left([c_{\ell q}^{(1)}]_{\alpha\alpha11}-[c^{(3)}_{\ell q}]_{\alpha\alpha11}-[c_{\ell d}]_{\alpha\alpha11}\right)\,,
    \label{eq:vector_nsi_matching_dA} 
\end{align}where $\delta g^Z$ and $\delta g^W$ contain the modified vertex between gauge bosons and fermions. Equivalently, the matching to the CC NSI in Eq.~\eqref{eq:CC-NSI} is given by~\cite{Falkowski:2017pss}:
\begin{align} 
%
%
\varepsilon^{\mu e L}_{\alpha\beta}&=-\dfrac{1}{2}[c_{\ell \ell}]_{\alpha\beta \mu e}+\delta_{\alpha\beta}[\delta g_L^{Ze}]_{\mu e}+\delta_{\mu \beta}[\delta g_L^{We}]_{e\alpha}+\delta_{e\alpha}[\delta g_L^{We}]_{\mu\beta}\,,\label{eq:matching_CCNSI_mueL}\\
\varepsilon^{\mu e R}_{\alpha\beta}&=-\dfrac{1}{2}[c_{\ell e}]_{\alpha\beta \mu e}+\delta_{\alpha\beta}[\delta g_R^{Ze}]_{\mu e}\,,\\
%
%
\varepsilon^{ud L}_{\alpha\beta}&=-[c_{\ell q}^{(3)}]_{\alpha\beta11}+[\delta g_L^{We}]_{\alpha\beta}\,,\label{eq:matching_CCNSI_udL}\\
\varepsilon^{ud R}_{\alpha\beta}&=0\,,\\
\varepsilon^{ud S}_{\alpha\beta}&=-\tfrac12\left([c_{\ell e qu}^{(1)}]^*_{\beta\alpha 11}+[c_{\ell e dq}]^*_{\beta\alpha 11}\right)\,,\label{eq:matching_CCNSI_udS}\\
\varepsilon^{ud P}_{\alpha\beta}&=-\tfrac12\left([c_{\ell e d q}]^*_{\beta\alpha 11}-[c_{\ell e  qu}^{(1)}]^*_{\beta\alpha 11}\right)\,,\\
\varepsilon^{ud T}_{\alpha\beta}&=-\tfrac12[c_{\ell e q u}^{(3)}]^*_{\beta\alpha 11}\,.
\label{eq:matching_CCNSI_udT}
\end{align}
 In these expressions, as we are considering only flavour violating CC NSI, $\alpha \neq \beta$ for the semileptonic operators while either $\alpha \neq e$ or $\beta \neq \mu$ for the fully leptonic ones\footnote{Notice that this implies no leading order corrections to the measurements of $G_F$ or $V_{ud}$ through $\mu$ and $\beta$ decays respectively and no contribution to $\varepsilon^{ud R}_{\alpha\beta}$ at $d=6$. Hence the simpler expressions reported.}.

\begin{table}[t!]
\renewcommand{\arraystretch}{1.51}
\begin{center}
    \begin{tabular}{c|c||c|c}
    \multicolumn{2}{c||}{up-quarks} & \multicolumn{2}{c}{down-quarks} \\
    \hline
    $\opd{\alpha \beta L}{uV} $  &  $(\bar u \gamma_\mu u)(\bar e_{L \alpha}  \gamma^\mu e_{L\beta}) $  &
    $\opd{\alpha \beta L}{dV} $  &  $(\bar d \gamma_\mu d)(\bar e_{L \alpha}  \gamma^\mu e_{L\beta}) $
    \\
    $\opd{\alpha \beta L}{uA} $   & $(\bar u \gamma_\mu \gamma_5 u)(\bar e_{L \alpha}  \gamma^\mu e_{L\beta})$   &
    $\opd{\alpha \beta L}{dA} $   & $(\bar d \gamma_\mu \gamma_5 d)(\bar e_{L \alpha}  \gamma^\mu e_{L\beta}) $
    \\
    $\opd{\alpha \beta R}{uV} $   & $(\bar u \gamma_\mu u)(\bar e_{R \alpha}  \gamma^\mu e_{R\beta})$  &
    $\opd{\alpha \beta R}{dV} $   & $(\bar d \gamma_\mu d)(\bar e_{R \alpha}  \gamma^\mu e_{R\beta}) $
    \\
    $\opd{\alpha \beta R}{uA} $   & $(\bar u \gamma_\mu \gamma_5 u)(\bar e_{R \alpha}  \gamma^\mu e_{R\beta})$ &
    $\opd{\alpha \beta R}{dA} $  & $ (\bar d \gamma_\mu \gamma_5 d)(\bar e_{R \alpha}  \gamma^\mu e_{R\beta}) $    
    \\[1ex]
    \hline
    $\opd{\alpha \beta R}{uS} $  & $ (\bar u  u)(\bar e_{L \alpha} e_{R\beta})   +\hc $ &
    $\opd{\alpha \beta R}{dS} $  & $ (\bar d  d)(\bar e_{L \alpha} e_{R\beta})   +\hc$
    \\
    $\opd{\alpha \beta R}{uP} $ & $(\bar u \gamma_5 u)(\bar e_{L \alpha}   e_{R\beta})   +\hc$ &
    $\opd{\alpha \beta R}{dP} $ & $(\bar d \gamma_5 d)(\bar e_{L \alpha}   e_{R\beta})   +\hc $
    \\
    $\opd{\alpha \beta R}{uT}$ &  $(\bar u  \sigma_{\mu \nu}  u)(\bar e_{L \alpha}   \sigma^{\mu \nu}   e_{R\beta})   +\hc $ &
    $\opd{\alpha \beta R}{dT} $&  $(\bar d  \sigma_{\mu \nu}  d)(\bar e_{L \alpha}   \sigma^{\mu \nu}   e_{R\beta})   +\hc$
    \\
    \end{tabular}
    \caption{List of relevant low-energy $d=6$ semileptonic operators involving charged lepton flavour change, some of them potentially related to neutrino NSI from the $SU(2)_L$ gauge invariance of the SMEFT framework.}
\label{tab:LEFT}
\end{center}
\end{table}

As it is manifest from the above expressions, the SMEFT generally induces at the same time both NC and CC NSI, a direct consequence of the $SU(2)_L$ gauge invariance. Furthermore, this also implies that the same SMEFT operators also generate at low energies operators with four charged fermions:
\begin{equation}
-\mathcal L_{\rm LEFT}^{\rm dim-6} \supset 
\frac1{v^2} \sum_{q,x,X}\, \wcL{\alpha\beta X}{qx}\,\mathcal O_{\alpha\beta X}^{qx}\,,
\end{equation}
where the corresponding operators are defined in Table~\ref{tab:LEFT}. From the following matching conditions to the operator basis defined in Ref.~\cite{Breso-Pla:2023tnz}, we indeed find that the same SMEFT operators leading to the LFV CC neutrino NSI listed above will also lead to LFV operators involving 2 charged leptons\footnote{Also 4-charged lepton operators will relate to the fully leptonic CC NSI. We have omitted this as they are constrained beyond the sensitivity of neutrino searches (see Ref.~\cite{Fernandez-Martinez:2024bxg}).}:
\begin{align}
%
c_{\alpha\beta L}^{u V} &= -\tfrac12 \left[c_{\ell u} + c^{(1)}_{\ell q} - c_{\ell q}^{(3)}\right]_{\alpha\beta 11}+\Big(1-\tfrac83\sw^2\Big) \left[\delta g_L^{Ze}\right]_{\alpha\beta},\label{eq:match_uVL}\\
c_{\alpha\beta L}^{u A} &= -\tfrac12 \left[c_{\ell u} - c^{(1)}_{\ell q} + c_{\ell q}^{(3)}\right]_{\alpha\beta 11}-\left[\delta g_L^{Ze}\right]_{\alpha\beta},\\
c_{\alpha\beta R}^{u V} &= -\tfrac12 \left[c_{e u} + c_{e q}\right]_{\alpha\beta 11}+\Big(1-\tfrac83\sw^2\Big) \left[\delta g_R^{Ze}\right]_{\alpha\beta},\\
c_{\alpha\beta R}^{u A} &= -\tfrac12 \left[c_{e u} - c_{e q}\right]_{\alpha\beta 11}-\left[\delta g_R^{Ze}\right]_{\alpha\beta},\\
c_{\alpha\beta L}^{d V} &= -\tfrac12 \left[c_{\ell d} + c^{(1)}_{\ell q} + c_{\ell q}^{(3)}\right]_{\alpha\beta 11}-\Big(1-\tfrac43\sw^2\Big) \left[\delta g_L^{Ze}\right]_{\alpha\beta},\\
c_{\alpha\beta L}^{d A} &= -\tfrac12 \left[c_{\ell d} - c^{(1)}_{\ell q} - c_{\ell q}^{(3)}\right]_{\alpha\beta 11}+\left[\delta g_L^{Ze}\right]_{\alpha\beta},\\
c_{\alpha\beta R}^{d V} &= -\tfrac12 \left[c_{e d} + c_{e q}\right]_{\alpha\beta 11}-\Big(1-\tfrac43\sw^2\Big) \left[\delta g_R^{Ze}\right]_{\alpha\beta},\\
c_{\alpha\beta R}^{d A} &= -\tfrac12 \left[c_{e d} - c_{e q}\right]_{\alpha\beta 11}+\left[\delta g_R^{Ze}\right]_{\alpha\beta},\label{eq:match_dAR}\\
c_{\alpha\beta R}^{uS}&=c_{\alpha\beta R}^{uP}=\tfrac12[c_{\ell e qu}^{(1)}]_{\alpha\beta 11},\label{eq:match_uSR}\\
c_{\alpha\beta R}^{dS}&=-c_{\alpha\beta R}^{dP}=-\tfrac12[c_{\ell e dq}]_{\alpha\beta 11},\\
c_{\alpha\beta R}^{uT}&=\tfrac12[c_{\ell e qu}^{(3)}]_{\alpha\beta 11},\\
c_{\alpha\beta R}^{dT}&=0\,.\label{eq:match_dT}
\end{align}
Given all these relations that the SMEFT induces at low energies, it is clear that a global analysis should consider not only those low-energy observables involving charged leptons, but also the neutrino sector by means of both CC and NC NSI. In this sense, some neutrino scattering data has been considered in previous SMEFT analyses~\cite{Falkowski:2017pss,Altmannshofer:2018xyo,Breso-Pla:2023tnz}, while the impact of reactor experiments has been analyzed in~\cite{Falkowski:2019xoe}, nevertheless a detailed study of the role of neutrino oscillation experiments is still missing. 

Performing a full general fit, considering all possible operators and observables, including all possible effects at neutrino oscillation experiments (production, oscillation and detection effects), is a very challenging task and beyond the scope of this work. Nevertheless, there are two interesting and consistent examples of how neutrinos can add information to the SMEFT analysis (and viceversa) that we present and study in the next sections.

\section{Global analysis of flavour conserving operators}
\label{sec:LFC}

In this section we focus on the lepton flavour conserving operators and study how the current constraints from neutrino experiments on the NSI, as globally derived in~\cite{Coloma:2023ixt}, can contribute to the global SMEFT picture. 
More specifically, our motivation is to understand if neutrino data can contribute and improve upon the global bounds on the SMEFT operators extracted from other low-energy observables, or even lifting some of the flat directions present. 

In order to address these questions, we start from the NSI constraints from the global SMEFT analysis in Ref.~\cite{Breso-Pla:2023tnz}, where a global fit to SMEFT operators was carried out by combining several sets of low-energy observables~\cite{CHARM:1986vuz,CHARM:1987pwr,Blondel:1989ev,CCFR:1997zzq,Qweak:2018tjf,Beise:2004py,Wood:1997zq,Edwards:1995zz,Vetter:1995vf,PVDIS:2014cmd,Argento:1982tq,UCNt:2021pcg,UCNA:2017obv,Markisch:2018ndu,CHARM:1988tlj,Ahrens:1990fp,CHARM-II:1994dzw,CCFR:1991lpl,CHARM-II:1990dvf,SLACE158:2005uay}, as well as collider data~\cite{Electroweak:2003ram,SLD:2000jop,ALEPH:2005ab,ALEPH:2013dgf,ABE1993288,VENUS:1997cjg,TOPAZ:2000evx,ATLAS:2017rue,Janot:2019oyi,dEnterria:2020cgt,D0:1999bqi,ATLAS:2016nqi,ATLAS:2020xea,LHCb:2016zpq,ATLAS:2018gqq,D0:2011baz,D0:2012kms,CDF:2012gpf,LHCb:2021bjt,ATLAS:2017rzl}. More specifically, we use their results, kindly provided by the authors of Ref.~\cite{Breso-Pla:2023tnz}, without including the neutrino-nucleus scattering data from COHERENT, since CE$\nu$NS data from both COHERENT~\cite{COHERENT:2017ipa,COHERENT:2018imc,COHERENT:2020iec} and Dresden-II~\cite{Colaresi:2021kus} was already included in the neutrino constraints on NSI of Ref.~\cite{Coloma:2023ixt}.

It is important to note that the results of~\cite{Coloma:2023ixt} were derived from CE$\nu$NS and oscillation data~\cite{Cleveland:1998nv,Kaether:2010ag,SAGE:2009eeu,Super-Kamiokande:2005wtt,Super-Kamiokande:2008ecj,Super-Kamiokande:2010tar,SK:nu2020,SNO:2005oxr,SNO:2006odc,SNO:2008gqy,SNO:2011hxd,Bellini:2011rx,Borexino:2008fkj,Borexino:2017rsf,KamLAND:2013rgu,DoubleC:nu2020,RENO:nu2020,DayaBay:2016ssb,DayaBay:2022orm,Wendell:2014dka,IceCube:2014flw,IceCube:2016rnb,MINOS:2013utc,T2K:nu2020,NOvA:nu2020} considering only NC NSI, while the SMEFT correlates CC and NC NSI and, in principle, both should be considered together in neutrino oscillation analyses. Thus, special care needs to be taken when combining them with the general constraints in the LFC SMEFT scenario from~\cite{Breso-Pla:2023tnz}. 
Nevertheless, there are two reasons that justify such a combination in this LFC scenario.

On the one hand, as argued in Section~\ref{sec:CC}, the effects of CC NSI often cancel when properly normalizing the probabilities in the LFC scenario. On the other hand, the observables considered in Ref.~\cite{Breso-Pla:2023tnz} impose constraints on CC NSI beyond the sensitivity of current neutrino oscillation experiments. 
In particular, the SMEFT operators that can lead to CC NSI and affect neutrino production and detection rates are vertex corrections as well as 4-fermion operators. The former are all bounded at the percent level or below by the global bounds derived in~\cite{Breso-Pla:2023tnz}, beyond current sensitivities of present oscillation experiments. Thus, the CC interactions that may potentially be probed through neutrino data would be:
\begin{equation}
    [c_{\ell q}^{(3)}]_{\alpha\alpha11}\,,[c_{\ell edq}]_{\alpha\alpha11}\,,[c_{\ell equ}^{(1)}]_{\alpha\alpha11}\,,[c_{\ell equ}^{(3)}]_{\alpha\alpha11}\,,\hspace{0.5cm}\text{for}\hspace{0.5cm}\alpha=e,\mu,\tau\,.
    \label{eq:CC_WCs}
\end{equation}
For operators involving electrons, all of the above operators are constrained at the percent level or below by $\beta$-decays, $\pi\rightarrow e\nu_e$ and LEP precision measurements~\cite{Breso-Pla:2023tnz}. Operators involving $\tau$'s are constrained by semileptonic $\tau$-decays. Moreover, they do not impact the observables considered in the neutrino NSI constraints of Ref.~\cite{Coloma:2023ixt}, as none of them involve the production or detection of neutrinos in association with a $\tau$. 

Regarding the muon sector, the vector operator $[c_{\ell q}^{(3)}]_{\mu\mu11}$ is constrained at the percent level by $\nu_\mu$-scattering on nucleons. Conversely, for the scalar/tensor operators two flat directions exist as only one direction is constrained by $\pi\rightarrow\mu\nu_\mu$. Since this decay is precisely how $\nu_\mu$ beams are produced, the flat directions do not play a role in production. In the detection process, scalar/tensor structures are chirality-suppressed, as they require a chirality flip with respect to the SM Lorentz structure, and their contribution can be neglected at the linear order approximation. 

Therefore, we conclude that it is consistent to neglect CC NSI in current neutrino data analysis when combined with the results of the LFC SMEFT fit of Ref.~\cite{Breso-Pla:2023tnz}, as their effects either cancel due to the normalisation of fluxes and cross sections, are constrained beyond neutrino data sensitivities by the SMEFT global fit, or only enter beyond the linear order. 

The SMEFT analysis is able to simultaneously constrain 65 WC combinations. Since not all individual WC entering the fit can be independently constrained, the flat directions are projected out by defining the following hatted coefficient combinations~\cite{Falkowski:2017pss,Breso-Pla:2023tnz}:
\begin{align}
        [\hat{c}_{eq}]_{ee11}&=[c_{e q}]_{ee11}+[c_{\ell q}^{(1)}]_{ee11}\,,\label{eq:adam_flat_directions1}\\
        [\hat{c}_{\ell u}]_{ee11}&=[c_{\ell u}]_{ee11}-[c_{eq}]_{ee11}\,,\label{eq:adam_flat_directions2}\\
        [\hat{c}_{\ell d}]_{ee11}&=[c_{\ell d}]_{ee11}-[c_{eq}]_{ee11}\,,\label{eq:adam_flat_directions3}\\
        [\hat{c}_{e u}]_{ee11}&=[c_{e u}]_{ee11}-[c_{\ell q}^{(1)}]_{ee11}\,,\label{eq:adam_flat_directions4}\\
        [\hat{c}_{e d}]_{ee11}&=[c_{e d}]_{ee11}-[c_{\ell q}^{(1)}]_{ee11}\,,\label{eq:adam_flat_directions5}\\
        [\hat{c}^{(3)}_{\ell q}]_{ee22}&=[c^{(3)}_{\ell q}]_{ee22}-[c_{\ell q}^{(1)}]_{ee22}\,,\label{eq:adam_flat_directions6}\\
        [\hat{c}_{\ell d}]_{ee22}&=[c_{\ell d}]_{ee22}+\left(5-3\dfrac{g^2}{g'^2}\right)[c_{\ell q}^{(1)}]_{ee22}-[\hat{c}_{eq}]_{ee11}\,,\label{eq:adam_flat_directions7}\\
        [\hat{c}_{e d}]_{ee22}&=[c_{e d}]_{ee22}+\left(3-3\dfrac{g^2}{g'^2}\right)[c_{\ell q}^{(1)}]_{ee22}-[\hat{c}_{eq}]_{ee11}\,,\label{eq:adam_flat_directions8}\\
        [\hat{c}^{(3)}_{\ell q}]_{ee33}&=[c^{(3)}_{\ell q}]_{ee33} + [c_{\ell q}^{(1)}]_{ee33}\,,\label{eq:adam_flat_directions9}\\
        [\hat{c}_{eq}]_{\mu\mu11}&=[c_{eq}]_{\mu\mu11}+[c_{ed}]_{\mu\mu11}-2[c_{eu}]_{\mu\mu11}\,,\label{eq:adam_flat_directions10}\\
        \epsilon_{P}^{d\mu}(2\text{ GeV})&=0.86[c_{\ell e dq}]_{\mu\mu11}-0.86[c_{\ell equ}]_{\mu\mu11}+0.012[c_{\ell equ}^{(3)}]_{\mu\mu11}\,,\label{eq:adam_flat_directions11}\\
        \epsilon_{P}^{d\tau}(2\text{ GeV})&=0.86[c_{\ell e dq}]_{\tau\tau11}-0.86[c_{\ell equ}]_{\tau\tau11}+0.012[c_{\ell equ}^{(3)}]_{\tau\tau11}\,,\label{eq:adam_flat_directions12}\\ 
        [\hat{c}_{\ell\ell}]_{\mu\mu\mu\mu}&=[c_{\ell\ell}]_{\mu\mu\mu\mu}+2\sw^2[c_{\ell e}]_{\mu\mu\mu\mu}\,.
    \label{eq:adam_flat_directions13}
\end{align}
With these definitions, a flat direction arises whenever the operator combination of the r.h.s.~conspires to cancel. 
It is then natural to wonder whether NSI constraints from neutrino oscillations data are sensitive to combinations of WC that may be able to lift any of these flat directions. 
This could be the case of the one associated to semileptonic operators involving the first leptonic and quark generations, Eqs.~\eqref{eq:adam_flat_directions1}-\eqref{eq:adam_flat_directions5}, since the rest involve operators to which neutrino oscillation experiments are not sensitive to. Unfortunately, neither the Earth propagation NSI $\varepsilon_{ee}^\oplus$, given in  Eq.~\eqref{eq:effective_Earth_NSI}, nor SNO's deuterium breakup (see Sec.~\ref{subsec:NC-NSI-Det}) are sensitive to this flat direction. This is a consequence of the fact that this direction is associated to an axial isoscalar combination of semileptonic operators, and $\varepsilon_{ee}^\oplus$ involves only vector operators, while SNO is only sensitive to the axial isovector NSI. 
An alternative way of seeing this is writing $\varepsilon_{ee}^{u,V}$ and $\varepsilon_{ee}^{d,V}$ in terms of the hatted coefficients:
\begin{align}
    \varepsilon_{ee}^{u,V}&=\delta g_L^{Zu} +\delta g_R^{Zu}+\left(1-\dfrac{8}{3}\sw^2\right)\delta g^{Z\nu_e}-\dfrac{1}{2}\left([c^{(3)}_{\ell q}]_{ee11}+[\hat{c}_{e q}]_{ee11}+[\hat{c}_{\ell u}]_{ee11}\right)\,,\label{eq:eps_uV_ee_hatted}\\
    \varepsilon_{ee}^{d,V}&=\delta g_L^{Zd} +\delta g_R^{Zd}-\left(1-\dfrac{4}{3}\sw^2\right)\delta g^{Z\nu_e}-\dfrac{1}{2}\left(-[c^{(3)}_{\ell q}]_{ee11}+[\hat{c}_{e q}]_{ee11}+[\hat{c}_{\ell d}]_{ee11}\right)\,,\label{eq:eps_dV_ee_hatted}
\end{align}
and equivalently for the axial isovector combination of NSIs entering SNO deuterium breakup cross section (see Ref.~\cite{Davidson:2003ha,Coloma:2023ixt}):
\begin{align}\label{eq:SNO}
\varepsilon^{\text{SNO}}& \approx \sum_\alpha\varepsilon_{\alpha\alpha}^{u,A}- \varepsilon_{\alpha\alpha}^{d,A} =3(\delta g_L^{Zu} -\delta g_L^{Zd})
 - 3(\delta g_R^{Zu}-\delta g_R^{Zd})
 +2\sum_\alpha\delta g^{Z\nu_\alpha}\nonumber\\
  &-\dfrac{1}{2}\left(2[ c^{(3)}_{\ell q}]+[\hat{c}_{\ell d}]-[\hat{c}_{\ell u}]\right)_{ee11}
  - \dfrac{1}{2}\left(2[ c^{(3)}_{\ell q}]+[c_{\ell d}]-[c_{\ell u}]\right)_{\mu\mu 11} \nonumber\\
  &- \dfrac{1}{2}\left(2[ c^{(3)}_{\ell q}]+[c_{\ell d}]-[c_{\ell u}]\right)_{\tau\tau 11}\,.
\end{align} 
Since each of the coefficients in Eqs.~\eqref{eq:eps_uV_ee_hatted} and~\eqref{eq:eps_dV_ee_hatted}, as well as the second row of Eq.~\eqref{eq:SNO}, is already bounded by the SMEFT analysis, a constraint on $\varepsilon_{ee}^\oplus$, $\varepsilon_{\mu\mu}^{\oplus}$ or $\varepsilon^{\text{SNO}}$ cannot close any of the blind directions presented above. Nevertheless, as we will show later, they can improve some of the already existing constraints on the SMEFT WC.

On the other hand, in the $\tau$ sector, $\varepsilon_{\tau\tau}^\oplus$ and SNO measurements do provide new independent information to the SMEFT global picture, as they probe operators that were not constrained at all before the inclusion of neutrino data and thus increase the number of constrained combinations.
All in all, the additional information from neutrino experiments that we will add on top of the already existing constraints is: three constraints on the LFC effective propagation NSI in Earth ($\varepsilon^{\oplus}_{ee}$, $\varepsilon^{\oplus}_{\mu\mu}$, $\varepsilon^{\oplus}_{\tau\tau}$), as well as a constraint on the axial NSI combination that modifies the deuterium breakup cross section in SNO.

\subsection{SMEFT impact on neutrino NSIs}

\begin{figure}[t!]
    \centering
    \includegraphics[width=\textwidth]{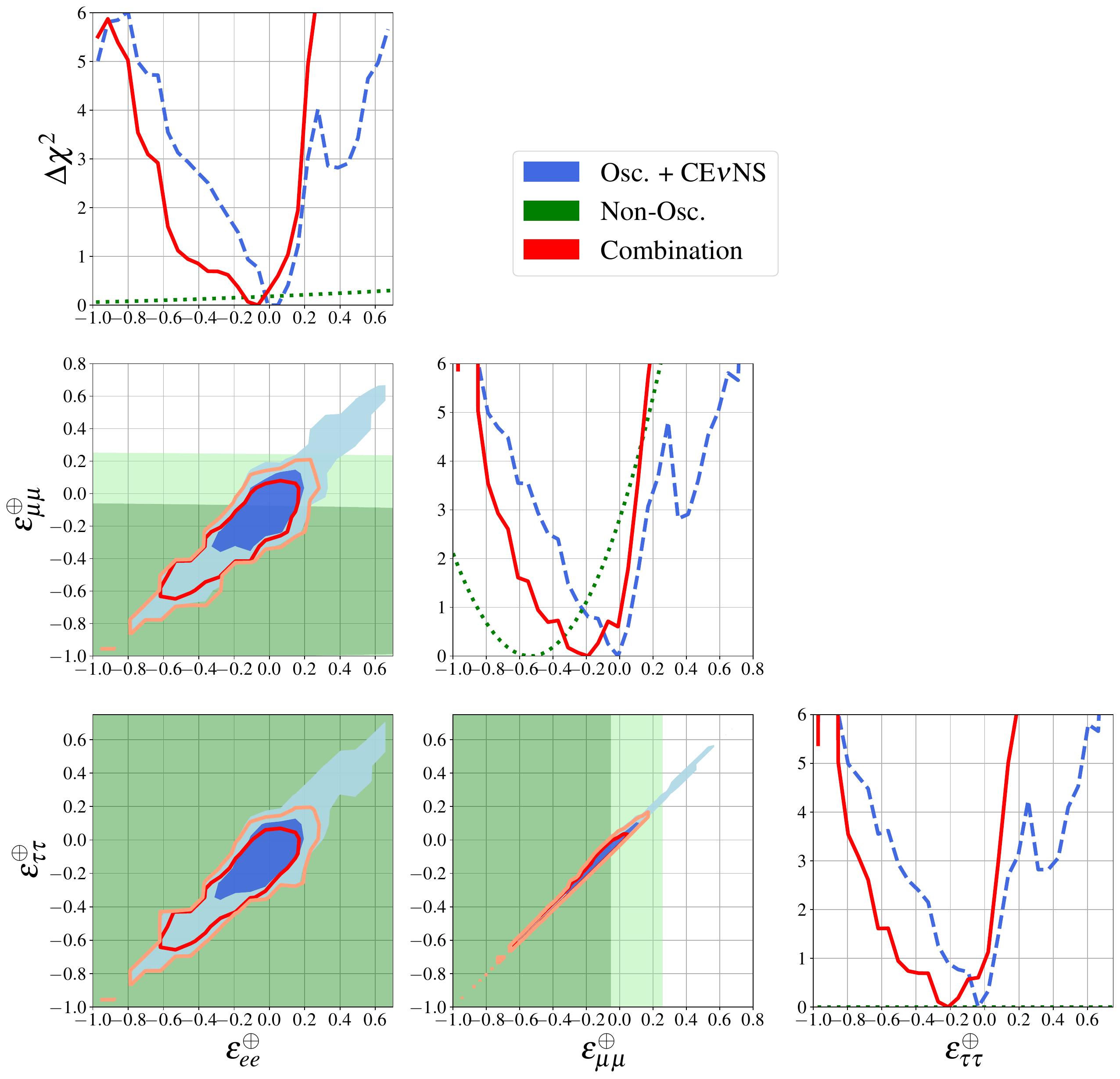}
    \caption{Correlations for the LFC effective propagation NSI in Earth $\varepsilon^{\oplus}_{\alpha\alpha}$. The bounds from the SMEFT analysis without including NSI data~\cite{Breso-Pla:2023tnz} are shown in green, the NSI contraints from neutrino data~\cite{Coloma:2023ixt} in blue, while our combination of the two is depicted in red. Darker (lighter) contours show the $1\sigma$ ($2\sigma$) allowed regions. }
    \label{fig:Earth_NSI_Combination}
\end{figure}

To show the role of the two datasets (NSI constraints from oscillations and CE$\nu$NS from~\cite{Coloma:2023ixt} and global constraints on SMEFT operators from~\cite{Breso-Pla:2023tnz}), we first project the results to the effective NSI parameters in Earth given in Eq.~\eqref{eq:effective_Earth_NSI}, and show also the bounds obtained from the combination of both sets of data in order to illustrate their complementarity. The results are shown in Fig.~\ref{fig:Earth_NSI_Combination}.

On the one hand, the global analysis of neutrino oscillation data and CE$\nu$NS provides similar sensitivities to all flavours, as well as strong correlations between them. The reason is that neutrino oscillation data alone is sensitive only to flavour differences $\varepsilon^{\oplus}_{\alpha\alpha}-\varepsilon^{\oplus}_{\beta\beta}$, and only the addition of CE$\nu$NS and neutrino-electron scattering allows to break this degeneracy~\cite{Coloma:2023ixt}. 

On the other hand, the global SMEFT analysis without the inclusion of oscillation and CE$\nu$NS data is not as stringent. 
It is not sensitive to $\varepsilon_{\tau\tau}^\oplus$ and the bound on $\varepsilon_{ee}^\oplus$ is extremely loose. The latter is a consequence of the fact that some of the SMEFT operators contributing to this NSI are only constrained by $\nu_e$-nucleus scattering in CHARM~\cite{CHARM:1986vuz} and, as a result, this very weak bound dominates the uncertainty in $\varepsilon_{ee}^\oplus$. 
The strongest constraint arises for $\varepsilon_{\mu\mu}^\oplus$, as $\nu_\mu$ scattering in nuclei is more precisely measured experimentally.
In fact, as Fig.~\ref{fig:Earth_NSI_Combination} shows, it is comparable to the one obtained in Ref.~\cite{Coloma:2023ixt} and the combination of both constraints yields a slightly ($\sim 20\%$) stronger bound.  Interestingly, this mild improvement also propagates to $\varepsilon_{ee}^\oplus$ and $\varepsilon_{\tau\tau}^\oplus$ through the strong correlations present between the NSI, as a consequence of the fact that their difference is more strongly constrained than their magnitude. This improvement, and the general results from the combined analysis, are summarized in Table~\ref{tab:NSIbounds}.

\begin{table}[t!]
\renewcommand{\arraystretch}{1.25}
\centering
    \begin{tabular}{|c|c|c|}
    \hline
    \multirow{2}{*}{NSI}&\multicolumn{2}{c|}{$95\%$ C.L. bound}\\\cline{2-3}
     &Osc. + CE$\nu$NS&Non-Osc. + Osc. + CE$\nu$NS\\\hline\hline
         $\varepsilon_{ee}^\oplus$&(-0.58, 0.50) & (-0.75, 0.16)\\\hline
         $\varepsilon_{\mu\mu}^\oplus$& (-0.61, 0.47)& (-0.79, 0.11)\\\hline
         $\varepsilon_{\tau\tau}^\oplus$ & (-0.62, 0.43) & (-0.80, 0.08)\\
         \hline\hline
         $\varepsilon_{ee}^\oplus-\varepsilon_{\mu\mu}^\oplus$ & (-0.14, 0.27) & (-0.14, 0.27)\\\hline
         $\varepsilon_{\tau\tau}^\oplus-\varepsilon_{\mu\mu}^\oplus$ & (-0.014, 0.025)&(-0.014, 0.025)\\\hline
           \end{tabular}
            \caption{Summary of the 95\% C.L. constraints on the NSI parameters in the LFC scenario obtained from neutrino data~\cite{Coloma:2023ixt} and our combined analysis also including the global SMEFT bounds from Ref.~\cite{Breso-Pla:2023tnz}. Note that, for the differences of the NSI coefficients, the inclusion of non-oscillation data does not lead to any improvement.}
            \label{tab:NSIbounds}
\end{table}

Figure~\ref{fig:Earth_NSI_Combination} illustrates the synergies between the SMEFT and neutrino oscillation programs, as combining NSI constraints from neutrino data with low-energy constraints within the SMEFT framework allows to strengthen the former. This also has the convenient side effect of smoothing the profiles leading to a more Gaussian behaviour. Indeed, we show in Appendix~\ref{App:gaussian_approximation}
that such approximation behaves well, and therefore we will use it in the next section to study the impact that NSI bounds from neutrino data has in the global SMEFT constraints. 

\subsection{Neutrino data impact on SMEFT}
\label{sec:NSIonSMEFT}

We focus next on how the global SMEFT picture is affected by the constraints on NSIs extracted from neutrino experiments and what kind of new information they provide to the general SMEFT program.
For this purpose, as before, we combine the likelihood from Ref.~\cite{Breso-Pla:2023tnz} not including COHERENT data with the NSI bounds of Ref.~\cite{Coloma:2023ixt} under the matching conditions of Eqs.~\eqref{eq:vector_nsi_matching_eV}-\eqref{eq:vector_nsi_matching_dA}.
In this case, however, the NSI bounds are included using the Gaussian approximation (see App.~\ref{App:gaussian_approximation}), as it simplifies the complex combination for such a large parameter space and allows us to provide combined correlation matrices as the final result. 

\begin{figure}[t!]
    \centering
    \includegraphics[width=\textwidth]{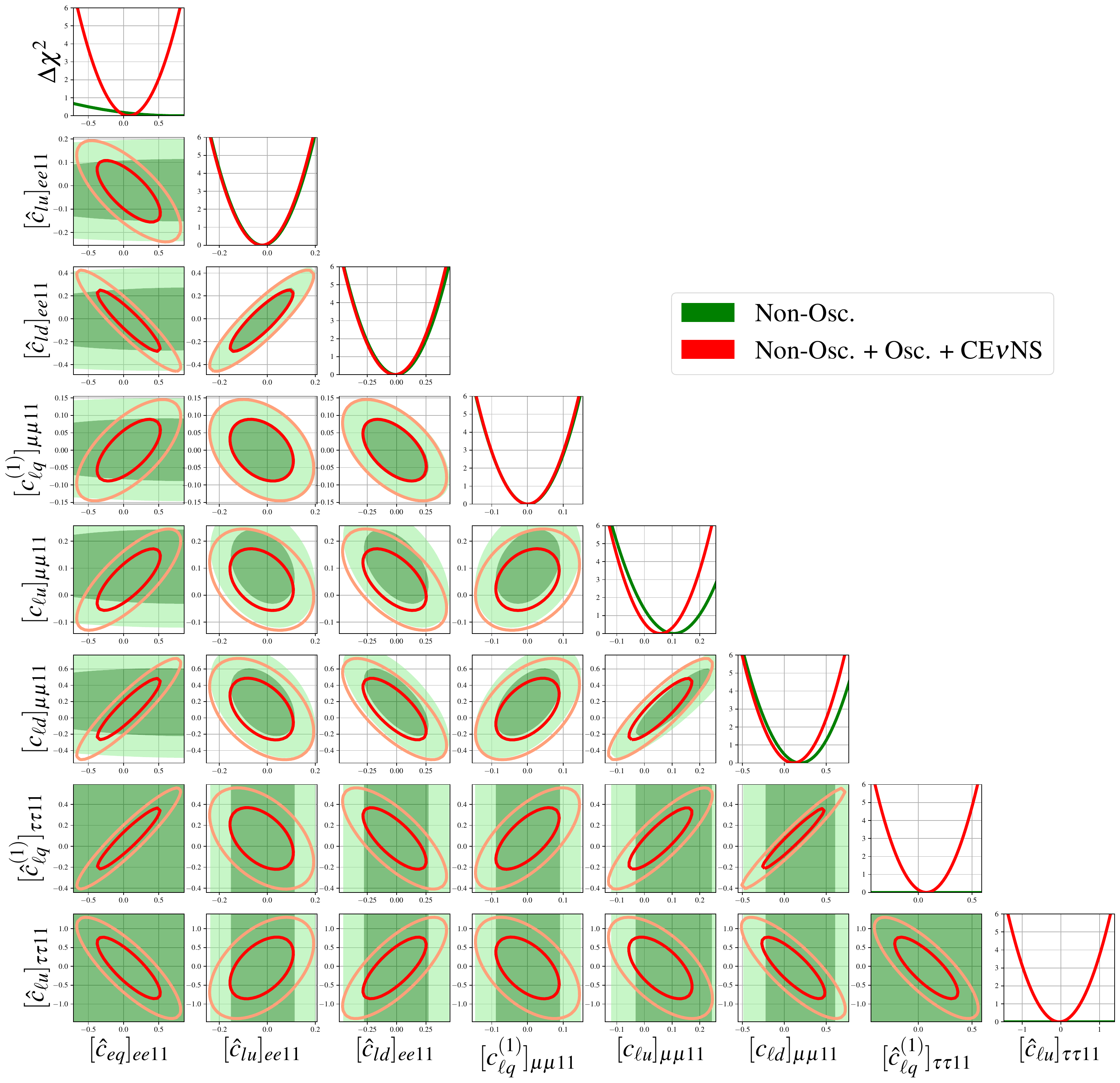}
    \caption{Correlations for the SMEFT operators that are affected by NSI constraints, with the same color code as Fig.~\ref{fig:Earth_NSI_Combination}. The addition of NSI bounds~\cite{Coloma:2023ixt} to the global SMEFT fit~\cite{Breso-Pla:2023tnz} not only improves the bounds for some operators, but also introduces new and strong correlations between them.}
    \label{fig:smeft_combination}
\end{figure}

Introducing oscillation and CE$\nu$NS data has several effects in the SMEFT global picture, as shown in Fig.~\ref{fig:smeft_combination}. 
Firstly, the weak bound on the combination of WC coefficients $[\hat{c}_{eq}]_{ee11}$, given by Eq.~(\ref{eq:adam_flat_directions1}) and mainly coming from CHARM, improves by a factor $\sim6$ with the inclusion of neutrino data, as previously noted in Ref.~\cite{Breso-Pla:2023tnz} with the inclusion of data from the COHERENT experiment.

Secondly, and more interestingly, the inclusion of neutrino data in the analysis allows for constraining NC semileptonic operators with the third lepton family, which previous SMEFT analyses were not sensitive to. 
The origin of this additional sensitivity is twofold. 
On the one hand, neutrino oscillations are sensitive to the difference of the effective NSI in propagation, $\varepsilon_{\alpha\alpha}^\oplus-\varepsilon_{\beta\beta}^\oplus$, and therefore they propagate the constraints on $\varepsilon_{ee}^\oplus$ and $\varepsilon_{\mu\mu}^\oplus$ (which involve more experimentally accessible operators) into $\varepsilon_{\tau\tau}^\oplus$, accessing operators that would be experimentally very challenging to probe directly at low energy. 
On the other hand, the combination of oscillations and CE$\nu$NS does set a direct constraint on $\varepsilon_{\tau\tau}^\oplus$, as well as the SNO bound in the deuterium breakup cross section, see Eq.~\eqref{eq:SNO}.
Translated to SMEFT coefficients, these two bounds define two new constrained combinations that were previously unbounded. 
\begin{align}
\label{eq:tauflatEarth}
    \varepsilon_{\tau\tau}^\oplus \longrightarrow [\hat{c}_{\ell q}^{(1)}]_{\tau\tau11}&=[c_{\ell q}^{(1)}]_{\tau\tau11}+\dfrac{2+Y_n^\oplus}{3(1+Y_n^{\oplus})}[c_{\ell u}]_{\tau\tau11}+\dfrac{1+2Y_n^\oplus}{3(1+Y_n^{\oplus})}[c_{\ell d}]_{\tau\tau11}\,,\\
    \label{eq:tauflatSNO}
    \text{SNO}\longrightarrow[\hat{c}_{\ell u}]_{\tau\tau11}&=[c_{\ell u}]_{\tau\tau11}-[c_{\ell d}]_{\tau\tau11}\,.
\end{align}
Note that the other operators contributing to SNO and $ \varepsilon_{\tau\tau}^\oplus$ are already bounded by other observables~\cite{Breso-Pla:2023tnz}.
These two equations define a new flat direction, which is a combination of WC previously unconstrained, to be added to the 9 directions already present in the SMEFT analysis~\cite{Falkowski:2017pss} without neutrino oscillations:
\begin{equation}
    \left(\textbf{F10}\right):\hspace{0.5cm}[c_{\ell u}]_{\tau\tau11}=[c_{\ell d}]_{\tau\tau11}=-[c_{\ell q}^{(1)}]_{\tau\tau11}\,.
\end{equation}
This flat direction could be lifted, for example, by measurements of the Drell-Yan di-tau differential cross section at the LHC~\cite{Allwicher_2023}.

Consequently, neutrino experiments do provide useful and complementary information to the global SMEFT picture, since they allow to improve the constraint on $[\hat{c}_{eq}]_{ee11}$, as previously outlined by Ref.~\cite{Breso-Pla:2023tnz} using COHERENT data. More interestingly, they provide constraints to the new operator combinations given by Eqs.~(\ref{eq:tauflatEarth})-(\ref{eq:tauflatSNO}).
Furthermore, it also generates correlations  where there were previously none, as can be seen in Fig.~\ref{fig:smeft_combination}.
For example, $[\hat{c}_{eq}]_{ee11}$ was uncorrelated with other semileptonic operators prior to the inclusion of oscillation data, while it is strongly correlated afterwards. 
This remarkable increase in the correlations within the coefficients in a global analysis implies that the global bounds on individual WC are more difficult to saturate, as precise cancellations between different operators are required.

We summarize our results in Table~\ref{tab:summary} with the constraints on the operators that are most significantly affected by the inclusion of neutrino oscillation and CE$\nu$NS data. Our results show that, neutrino experiments, including neutrino oscillation data, provide important information for the SMEFT program and should be considered in global analyses.

\begin{table}
\renewcommand{\arraystretch}{1.15}
\centering
\begin{tabular}{|c|c|c|}
\hline
     \multirow{2}{*}{Operators}&\multicolumn{2}{c|}{1$\sigma$ interval}\\\cline{2-3}
     &Non-Osc.& Non-Osc. + Osc. + CE$\nu$NS\\\hline\hline
     $[\hat{c}_{eq}]_{ee11}$&$0.76\pm 1.80$ & $0.07\pm0.30$\\\hline
     $[c_{\ell u}]_{\mu\mu11}$&$0.110\pm 0.091$&$0.058 \pm 0.076$\\\hline
     $[c_{\ell d}]_{\mu\mu11}$&$0.19\pm 0.27$&$0.11 \pm 0.25$\\\hline
     $[\hat{c}_{\ell q}^{(1)}]_{\tau\tau11}$ & Unconstrained  & $0.07\pm0.19$\\\hline
     $[\hat{c}_{\ell u}]_{\tau\tau11}$ & Unconstrained  & $-0.04\pm 0.54$\\\hline
\end{tabular}
\caption{Summary of the largest improvements for the global bounds for SMEFT operators after the addition of information from NSI. Notice, however, that the latter also impact the global analysis introducing new correlations between operators, as seen in Fig.~\ref{fig:smeft_combination}. 
} \label{tab:summary}      
\end{table}

\section{Global analysis of flavour violating operators}
\label{sec:LFV}

\begin{figure}[t!]
    \centering
    \includegraphics[width=0.7\textwidth]{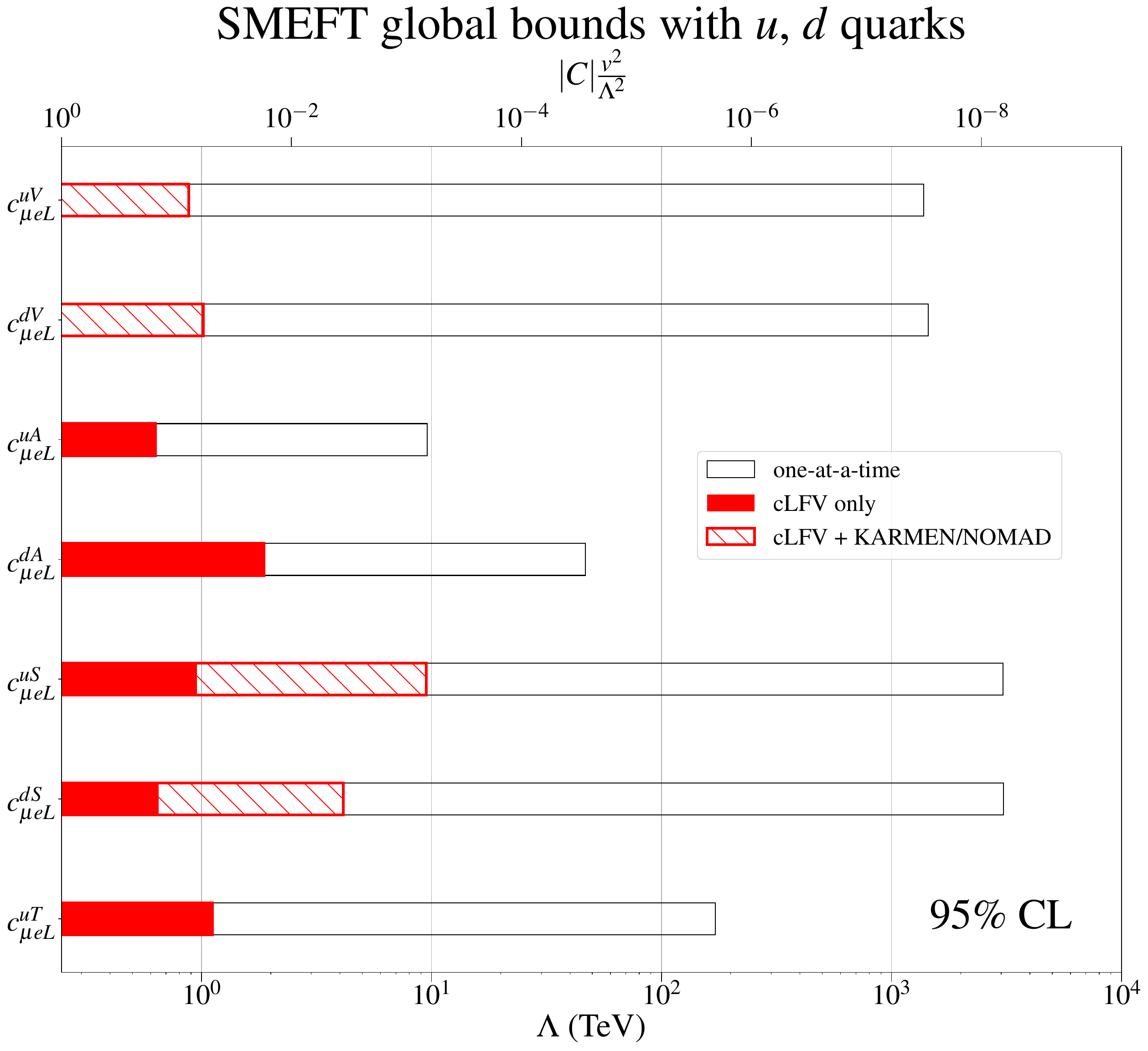}
    \caption{Current 95\% bounds on the $\mu-e$ LFV low-energy operators, as induced by dim=$6$ SMEFT, involving only first generation quarks and left-handed electrons. 
    The empty bars correspond to the bounds extracted considering the presence of one operator at a time.
    The solid bars correspond to the global bounds of~\cite{Fernandez-Martinez:2024bxg}, derived considering all operators at the same time but including only cLFV observables. The hatched bars represent the improvement of these global bounds upon the addition of the zero-distance neutrino oscillation constraints from KARMEN and NOMAD.}
    \label{fig:cLFV_global_L_bounds}
\end{figure}

\begin{figure}[t!]
    \centering
    \includegraphics[width=0.7\textwidth]{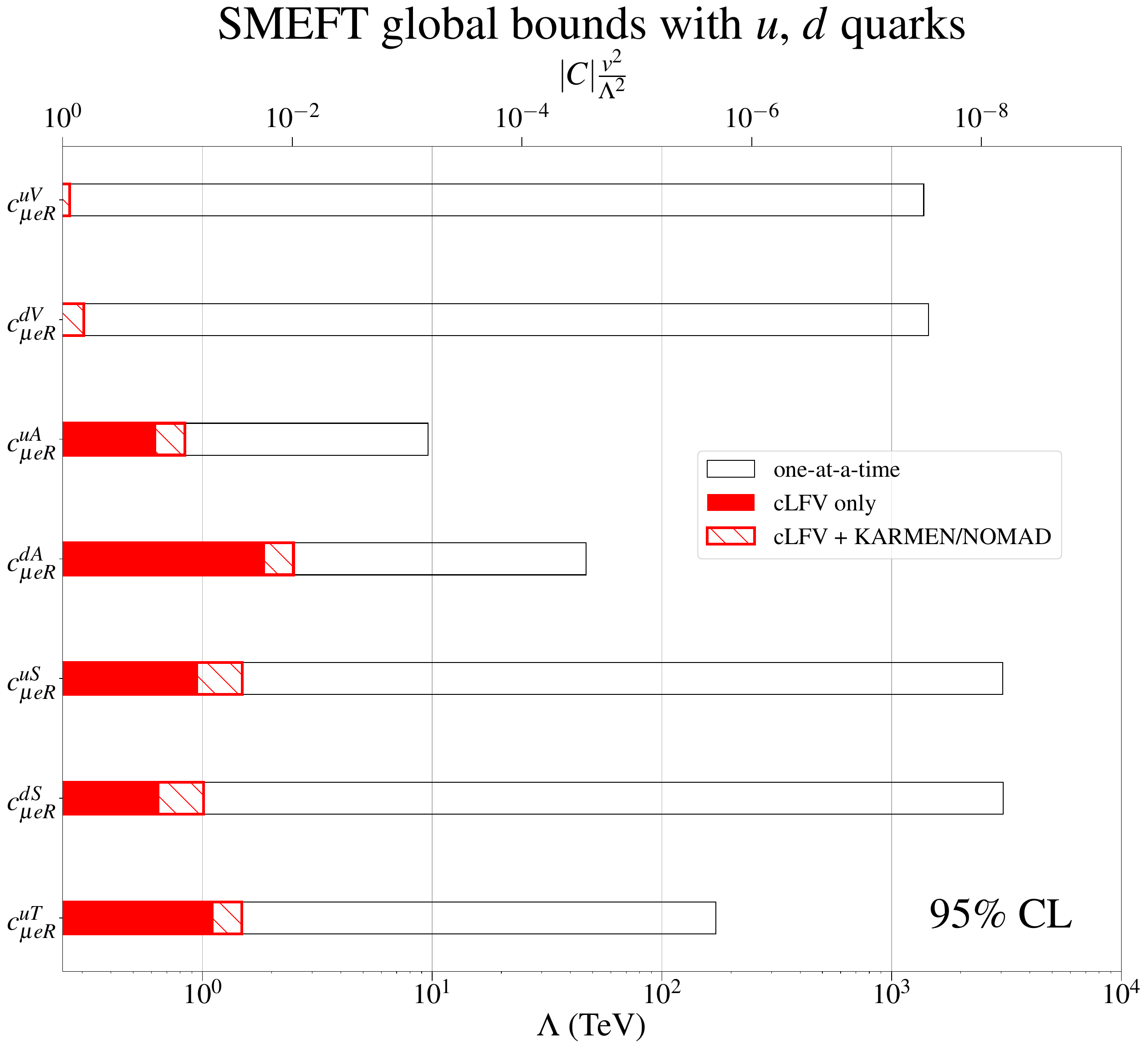}
    \caption{Same as Fig.~\ref{fig:cLFV_global_L_bounds} but for operators with right-handed electrons.}
    \label{fig:cLFV_global_R_bounds}
\end{figure}

Recently, a global analysis of low-energy charged LFV (cLFV) observables~\cite{Badertscher:1981ay,SINDRUM:1987nra,SINDRUMII:1993gxf,White:1995jc,SINDRUMII:1996fti,CLEO:1999nsy,Appel:2000wg,SINDRUMII:2006dvw,BaBar:2006jhm,KTeV:2007cvy,Belle:2007cio,Achasov:2009en,BaBar:2009hkt,Hayasaka:2010np,Belle:2012unr,NA62:2021zxl,Belle:2021ysv,MEGII:2023ltw,Belle:2023ziz} was carried out in~\cite{Fernandez-Martinez:2024bxg}, identifying the presence of poorly-constrained directions involving semileptonic operators. In the case of the $d=6$ SMEFT, the plethora of bounds on different LFV semileptonic $\tau$ decays offer great complementarity so as to efficiently constrain the different directions in WC space.
Conversely, the $\mu-e$ sector parameter space has some directions very stringently constrained but rather loose bounds in others, allowing to relax the global bounds by many orders of magnitude with respect to those extracted by considering one operator at a time, when fine tuned cancellations are permitted. This situation is summarized through Figs.~\ref{fig:cLFV_global_L_bounds} and~\ref{fig:cLFV_global_R_bounds} where, for the different operators, the empty bands reflect the present bound when only one operator at a time is considered (and hence no cancellations are allowed), while the filled red bands show how these constraints are relaxed in the context of the global fit from Ref.~\cite{Fernandez-Martinez:2024bxg} due to the poorly constrained directions.
It is therefore interesting to consider whether neutrino experiments may improve these global results. 

Neutrinos do indeed provide information about charged lepton operators exploiting their relation through $SU(2)_L$ gauge invariance, which in the SMEFT correlates both, as can be explicitly seen from the matching conditions in Eqs.~\eqref{eq:matching_CCNSI_mueL}-\eqref{eq:matching_CCNSI_udT} and \eqref{eq:match_uVL}-\eqref{eq:match_dT}. Contrary to the LFC case, there are no compelling arguments to neglect the impact of CC LFV NSI in a global neutrino data fit in the SMEFT context. Therefore, it is not consistent to combine the global fit to NC NSI given in~\cite{Coloma:2023ixt} with the cLFV SMEFT analysis given in~\cite{Fernandez-Martinez:2024bxg}. Nevertheless, a very effective way of constraining CC LFV NSI is through the zero distance effect described by Eq.~\eqref{eq:correctzerodistap}. In particular, KARMEN~\cite{Eitel:2000by} and NOMAD~\cite{NOMAD:2001xxt,NOMAD:2003mqg} provide excellent probes of this effect in the $\mu-e$ sector:
\begin{align}
\tilde{P}^{\text{KARMEN}}_{\bar{\nu}_\mu\rightarrow \bar{\nu}_e}&=
\sum_\alpha p^{\mu}_{XY}\nsi{\mu e X}{\alpha e}\nsi{\mu eY}{\alpha e}+d^{\beta}_{XY}\nsi{ud X}{e\mu}\nsi{udY}{e\mu} + 2d^{\beta}_{XL}p^{\mu}_{YL} \varepsilon_{e\mu}^{ud X} \varepsilon_{e e}^{\mu e Y}
<6.5\cdot 10^{-4}\,,\\
\tilde{P}^{\text{NOMAD}}_{\nu_\mu\rightarrow\nu_e}&=
p^{\pi}_{XY}\nsi{udX}{\mu e}\nsi{udY}{\mu e}+d^{ud}_{XY}\nsi{udX}{e\mu}\nsi{udY}{e\mu}{+2d^{ud}_{XL}p^{\pi}_{YL}\nsi{udX}{e\mu}\nsi{udY}{\mu e}}
<7.0\cdot 10^{-4}\,,
\end{align}
at the $90\%$ CL. Here $p^S_{XY}$ and $d^P_{XY}$ are, respectively, production and detection coefficients, computed for some neutrino source $S$ and detection process $P$, and summation over repeated $X,Y=L,R,S,P,T$ is understood. The relevant production/detection coefficients are~\cite{Falkowski:2019kfn,Falkowski:2021bkq}: 
\begin{align}
    {\rm KARMEN} & 
     \begin{cases} 
      p_{LL}^\mu=1,\hspace{0.5cm}p_{RR}^\mu=\dfrac{6m_\mu - 12 E_\nu}{3 m_\mu - 4E_\nu},\\
     \\
     d^\beta_{LL}=1,\hspace{0.5cm}d_{SS}^\beta=\dfrac{g_S^2}{1+3g_A^2},\hspace{0.5cm}d_{TT}^\beta=\dfrac{g_T^2}{1+3g_A^2},
   \end{cases}
\\\nonumber\\
    {\rm NOMAD} & 
    \begin{cases} 
      p_{LL}^\pi=1,\hspace{0.5cm}
      p_{PL}^\pi=\dfrac{m_\pi^2}{m_\mu(m_u+m_d)},\hspace{0.5cm}p_{PP}^\pi=\dfrac{m_\pi^4}{m_\mu^2(m_u+m_d)^2},\\
     \\
     d^{ud}_{LL}\simeq0.91,\hspace{0.5cm}d_{SS}^{ud}=d_{PP}^{ud}\simeq0.04,\hspace{0.5cm}d_{TT}^{ud}\simeq0.59,
   \end{cases}
\end{align}
where KARMEN's detection coefficients have been computed in the inverse-$\beta$ regime neglecting nucleon recoil, and $g_A$, $g_S$ and $g_T$ are, respectively, the axial, scalar and tensor charges of the nucleon. As for NOMAD, detection coefficients refer to the DIS regime extracted from Ref.~\cite{Falkowski:2021bkq}. In both cases we have not shown the detection coefficients involving the operators with right-handed quark currents as they are either flavour conserving or are higher order when matching to SMEFT operators~\cite{Jenkins:2017jig}.  
Production via $\mu$ and $\pi$ decay has been assumed for KARMEN and NOMAD, respectively. We have also neglected the electron mass and, as such, the interferences between operators with different electron chirality are not present. Notice that the normalization effects discussed in Section~\ref{sec:CC} are subleading in appearance channels. 

In Figs.~\ref{fig:cLFV_global_L_bounds} and~\ref{fig:cLFV_global_R_bounds} we show the improvement upon including the constraints from KARMEN and NOMAD in the global fit to cLFV in the $\mu-e$ sector from Ref.~\cite{Fernandez-Martinez:2024bxg} by the hatched red bars\footnote{Here we restrict to the simplest SMEFT scenario of~\cite{Fernandez-Martinez:2024bxg} with only first generation quarks. The discussion is qualitatively the same for the other SMEFT scenario in~\cite{Fernandez-Martinez:2024bxg} also with $s$ quarks.}. It is straightforward to translate the bounds from KARMEN and NOMAD on scalar, pseudo-scalar and tensor CC NSI into constraints on the semileptonic cLFV scalar and tensor operators, see Eqs.~(\ref{eq:matching_CCNSI_udS})-(\ref{eq:matching_CCNSI_udT}) and Eqs.~(\ref{eq:match_uSR})-(\ref{eq:match_dT}). However, the correlations among CC and NC interactions induced by the vectorial operator $[c_{\ell q}^{(3)}]$ are less straightforward, as the matching to vectorial low energy 4-fermion operators also involves the vertex corrections with the $W$ and $Z$ bosons. For these Lorentz structures, translating the neutrino bounds to the charged lepton sector is more convoluted. The dependence on the SMEFT Wilson Coefficients of the CC-NSI relevant for production and detection in KARMEN and NOMAD is summarized in Table~\ref{tab:my_label} (those in grey are more strongly bounded by non-neutrino data as discussed below).

The cLFV Z-vertex corrections induce LFV Z boson decays~\cite{Crivellin:2013hpa,Calibbi:2021pyh}, which are stringlently constrainted by LHC~\cite{ATLAS:2022uhq,ATLAS:2021bdj}. These translate into much stronger constraints on $[\delta g^{Ze}]_{\alpha\beta}$ than the global bounds on cLFV operators from Ref.~\cite{Fernandez-Martinez:2024bxg}, so their contribution can be neglected. Similarly, 4-lepton operators potentially contributing to production via $\mu$ decay at KARMEN, may be neglected given the stronger constraints from LFV leptonic decays~\cite{Fernandez-Martinez:2024bxg}.
On the other hand, the LFV W-vertex corrections are quite loosely constrained, since the neutrino flavour is not observed. 
Interestingly, it is precisely experiments such as KARMEN/NOMAD that are able to observe the flavour in CC interactions, potentially allowing to disentangle the W-vertex correction from the 4-fermion operator $c_{\ell q}^{(3)}$.
In fact, KARMEN and NOMAD are sensitive to different combinations of the W-vertex $[\delta g_L^W]_{\mu e}$ and 4-fermion operator $c_{\ell q}^{(3)}$ contributions thanks to their different production channels ($\mu$ vs $\pi$ decay). As such, the combination of the two bounds is very complementary and allows to constrain both operators independently. Together with the bounds on the scalar and tensor structures appearing in the detection processes and in production via $\pi$ decay this leads to the improvement shown in Figs.~\ref{fig:cLFV_global_L_bounds} and~\ref{fig:cLFV_global_R_bounds}.

Since $[c_{\ell q}^{(3)}]_{e\mu 11}$ enters in the matching of left-handed cLFV vector and axial operators with $u$ and $d$ quarks, which were only loosely bounded above $\mathcal O(1)$ in the global fit of Ref.~\cite{Fernandez-Martinez:2024bxg}, the bound derived from KARMEN and NOMAD improves the global bound on vector operators by an order of magnitude, as it can be seen in Fig.~\ref{fig:cLFV_global_L_bounds}. This improvement also propagates to scalar operators, due to the strong correlation with vector operators induced by the experimental constraints as shown in~\cite{Fernandez-Martinez:2024bxg}. 
Unfortunately, this improvement does not affect axial operators, as the poorly constrained direction that relaxes the global bound corresponds to the isoscalar $u+d$ combinations, while $c_{\ell q}^{(3)}$ only contributes to the $u-d$ combination.

The situation is much different for operators in which the leptonic current is right-handed, as their $SU(2)_L$ singlet nature does not correlate charged leptons and neutrinos for vector operators. As a consequence, the only correlation present in this sector between neutrinos and charged leptons is associated to the scalar/tensor operators, as they are, in this case, the only ones receiving contributions from operators containing $SU(2)_L$ doublets. The results are shown in Fig.~\ref{fig:cLFV_global_R_bounds}. Since scalar and tensor operators are better constrained than vectors, the improvement obtained by adding KARMEN/NOMAD is not as noticeable as for the left-handed case, but still shows that neutrino experiments do contribute improving the bounds in the global SMEFT picture. The marginal improvement in the vector and and axial structures is indirect since, from the global fit to cLFV observables, these operators were correlated with the scalar and tensor ones respectively~\cite{Fernandez-Martinez:2024bxg}.

\begin{table}[t!]
\renewcommand{\arraystretch}{1.3}
    \centering
    \begin{tabular}{|c|c|c|}
    \hline
    Process&Low-energy WC &SMEFT WC\\
    \hline\hline
       KARMEN& $\varepsilon_{ee}^{\mu eL}$, ${\color{gray}\varepsilon_{\mu e}^{\mu eL}}$, ${\color{gray}\varepsilon_{\tau e}^{\mu eL}}$ & ${\color{gray}{[c_{\ell \ell}]_{\alpha e\mu e}\color{gray}}}$, ${\color{gray}[\delta g_L^{Ze}]_{\mu e}}$, $[\delta g_L^{We}]_{\mu e}$ \\
        production& ${\color{gray}\varepsilon_{ee}^{\mu eR}}$, ${\color{gray}\varepsilon_{\mu e}^{\mu eR}}$, ${\color{gray}\varepsilon_{\tau e}^{\mu eR}}$&  ${\color{gray}[c_{\ell e}]_{\alpha e\mu e}}$, ${\color{gray}[\delta g_R^{Ze}]_{\mu e}}$\\
    \hline
        KARMEN & \multirow{2}{*}{$\varepsilon^{ud L}_{e\mu}$, $\varepsilon^{udS}_{e\mu}$, $\varepsilon_{e\mu}^{udT}$}& $[c_{\ell q}^{(3)}]_{e\mu 11}$, $[\delta g_L^{We}]_{e\mu}$\\
        detection & &$[c_{\ell e dq}]^*_{\mu e11}$, $[c_{\ell e qu}^{(1)}]^*_{\mu e11}$, $[c_{\ell e qu}^{(3)}]^*_{\mu e11}$\\
    \hline\hline
    NOMAD & \multirow{2}{*}{$\varepsilon_{\mu e}^{udL}$, $\varepsilon_{\mu e}^{udP}$}& $[c_{\ell q}^{(3)}]_{\mu e 11}$, $[\delta g^{We}_L]_{\mu e}$ \\
    production& &$[c_{\ell e dq}]^*_{e\mu11}$, $[c_{\ell e qu}^{(1)}]^*_{e\mu11}$\\
    \hline
    NOMAD &$\varepsilon^{ud L}_{e\mu}$, $\varepsilon^{ud S}_{e\mu}$ & $[c_{\ell q}^{(3)}]_{e\mu 11}$, $[\delta g_L^{We}]_{e\mu}$\\
    detection& $\varepsilon^{ud P}_{e\mu}$, $\varepsilon^{ud T}_{e\mu}$&$[c_{\ell e dq}]^*_{\mu e11}$, $[c_{\ell e qu}^{(1)}]^*_{\mu e11}$, $[c_{\ell e qu}^{(3)}]^*_{\mu e11}$\\
    \hline
    \end{tabular}
    \caption{List of low-energy and SMEFT Wilson coefficients relevant for production and detection in KARMEN and NOMAD. Those in grey are already (globally) bounded by cLFV observables~\cite{Fernandez-Martinez:2024bxg} beyond the sensitivity of neutrino experiments.}
    \label{tab:my_label}
\end{table}

\section{Conclusions}
\label{sec:Conclusions}

The $SU(2)_L$ gauge invariance necessarily links new physics involving the neutrino and charged lepton sectors at some level. Given that the charged lepton sector is much more easily probed and constrained, neutrino non-standard interactions (NSI) are expected to be bounded beyond the sensitivity of present and near-future neutrino experiments~\cite{Antusch:2008tz,Gavela:2008ra} when induced by new heavy mediators and barring fine-tuned cancellations. Nevertheless, global constraints on the SMEFT operators have matured and reached a level of precision where the role of unconstrained, flat directions is being actively investigated with effort in identifying complementary observables able to lift them and make more substantial contributions to the global picture. 

In this context, we have investigated the connection between the NSI formalism and the Standard Model effective field theory (SMEFT), focusing on how neutrino data contribute to the global constraints on the latter. Building upon the  analyses of Refs.~\cite{Falkowski:2019kfn,Breso-Pla:2023tnz,Cherchiglia:2023aqp}, we have first presented the SMEFT operators that contribute to neutrino NSI when the corresponding mediators are heavier than the electroweak scale. We have also emphasized how the measurements of neutrino fluxes and cross sections required to extract the oscillation probabilities in neutrino experiments are also affected by this type of new physics. In particular, we have clarified how this can lead to indirect contributions that tend to cancel the sensitivity of these experiments to charged current (CC) flavour-conserving NSI. 

We have also investigated the connection between neutrino NSI and charged lepton flavour violating (cLFV) operators implied by the $SU(2)_L$ gauge invariance built into the SMEFT. This same gauge invariance generally relates CC and NC NSI. As such, to derive the most comprehensive constraints on these effective operators, a global analysis of neutrino oscillation and scattering data together with relevant observables involving charged leptons would be necessary. In this work we have performed the first steps towards this ambitious goal studying the impact of neutrino data in two meaningful simplified scenarios.

First, we have made the common simplifying assumption of flavour conservation and started from the comprehensive global fit to flavour-conserving SMEFT operators from LEP and low-energy observables presented in Refs.~\cite{Falkowski:2017pss,Breso-Pla:2023tnz}. Since in this scenario the sensitivity to CC NSI from neutrino oscillation data is very limited from the indirect effects stemming from the flux and cross section measurements and the strong constraints from the SMEFT global fit itself, it is sufficient to consider only NC NSI when analyzing neutrino data. Thus, we have combined the results from the global fit of~\cite{Breso-Pla:2023tnz} (without the inclusion of COHERENT data) with those on NC neutrino NSI from neutrino oscillation and CE$\nu$NS data from Ref.~\cite{Coloma:2023ixt}. In both analyses the impact of LFV operators was marginalized thus providing conservative constraints to a LFC scenario.

We find that, as observed in~\cite{Breso-Pla:2023tnz}, upon the inclusion of CE$\nu$NS data, there is an improvement of about an order of magnitude on the bound of an operator contributing to electron and $\nu_e$ scattering with quarks. Moreover, we find that, given the stringent correlations among different flavours in the neutrino NSI constraints from oscillation data, this improvement propagates also to the $\mu$ and $\tau$ sectors. This, together with data on NC neutrino detection from SNO, allows to derive constraints on two operator combinations involving the $\tau$ flavour that were previously unconstrained in the analysis of Ref.~\cite{Breso-Pla:2023tnz}. We also find that, upon the inclusion of the SMEFT constraints in the NSI analysis, the bounds on SMEFT-generated NSI involving $\nu_\mu$ strengthen. Interestingly, the correlations of NSI constraints with other flavours lead to indirect improvements in the $e$ and $\tau$ sectors as well. 

For the second scenario, we concentrate instead on LFV operators. Recently, Ref.~\cite{Fernandez-Martinez:2024bxg} derived global constraints on LFV SMEFT operators from charged lepton data. Their results show that, while the different bounds on LFV semileptonic $\tau$ decays allow to derive meaningful global constraints on all the relevant operators of the $\tau-e$ and $\tau-\mu$ sectors, in the $e-\mu$ sector several poorly constrained directions exist. This leads to surprisingly weak global bounds on some operator combinations involving $e-\mu$ LFV. Here, we have studied how several of these poorly-constrained directions are improved by about an order of magnitude with constraints from KARMEN and NOMAD on short-baseline $\nu_e$ appearance. 

Our results show that neutrino data, both from oscillation and scattering measurements, provide significant contributions to the global SMEFT program. They both lead to improvements by even an order of magnitude in some operators and provide constraints on previously unbounded directions involving the more challenging $\tau$ flavour. Thus, neutrino data turn out to be very complementary and valuable and it is worth to incorporate it to global SMEFT analyses.


\paragraph{Acknowledgments.}
We thank Michele Maltoni, Concha Gonzalez-Garcia and João Paulo Pinheiro for kindly providing their neutrino oscillation and CE$\nu$NS fit results. We also warmly thank Martín González-Alonso for sharing the SMEFT likelihood. This project has received support from the European Union’s Horizon 2020 research and innovation programme under the Marie Skłodowska-Curie grant agreement No~860881-HIDDeN and No 101086085 - ASYMMETRY, and from the Spanish Research Agency (Agencia Estatal de Investigaci\'on) through the Grant IFIC Centro de Excelencia Severo Ochoa No CEX2023-001292-S, Grant ID2020-113644GB-I00, Grant IFT Centro de Excelencia Severo Ochoa No CEX2020-001007-S and Grant PID2019-108892RB-I00 funded by MCIN/AEI/10.13039/501100011033. XM acknowledges funding from the European Union’s Horizon Europe Programme under the Marie Skłodowska-Curie grant agreement no.~101066105-PheNUmenal. 
The work of DNT was supported by the Spanish MIU through the National Program FPU (grant number FPU20/05333). JLP also acknowledges financial support from Generalitat Valenciana through the plan GenT program (CIDEGENT/2018/019), from the Spanish Research Agency (Agencia Estatal de Investigaci\'on) through grant CNS2022-136013 funded by MICIU/AEI/10.13039/501100011033 and by “European Union NextGenerationEU/PRTR'', and from the MCIU with funding from the European Union NextGenerationEU (PRTR-C17.I01) and Generalitat
Valenciana (ASFAE/2022/020). PC acknowledges support from Grant PID2022-142545NB-C21 funded by MCIN/AEI/10.13039/501100011033/ FEDER, UE, from the Spanish Research Agency through the grant CNS2023-145338 funded by MCIN/AEI/10.13039/501100011033 and by “European Union NextGenerationEU/PRTR''. She is also supported by grant RYC2018-024240-I, funded by MCIN/AEI/10.13039/501100011033 and by ``ESF Investing in your future''.

\appendix

\section{Gaussian approximation}
\label{App:gaussian_approximation}

As it can be seen from Ref.~\cite{Coloma:2023ixt}, the $\Delta\chi^2$ profiles for the $\varepsilon_{\alpha\alpha}^{\oplus}$ are very non-Gaussian. However, upon the combination of these results with the SMEFT bounds from Ref.~\cite{Breso-Pla:2023tnz} they acquire a more Gaussian profile. In this Appendix, we explore whether the aforementioned combination is reasonably well approximated by a Gaussian distribution. This will greatly simplify the exploration of the vast SMEFT parameter space to investigate the impact of the neutrino data on it. The final results of the combination can also be presented in a compact and comprehensive form through the corresponding combined covariance matrix. 

\begin{figure}[t!]
    \centering
    \includegraphics[width=.85\textwidth]{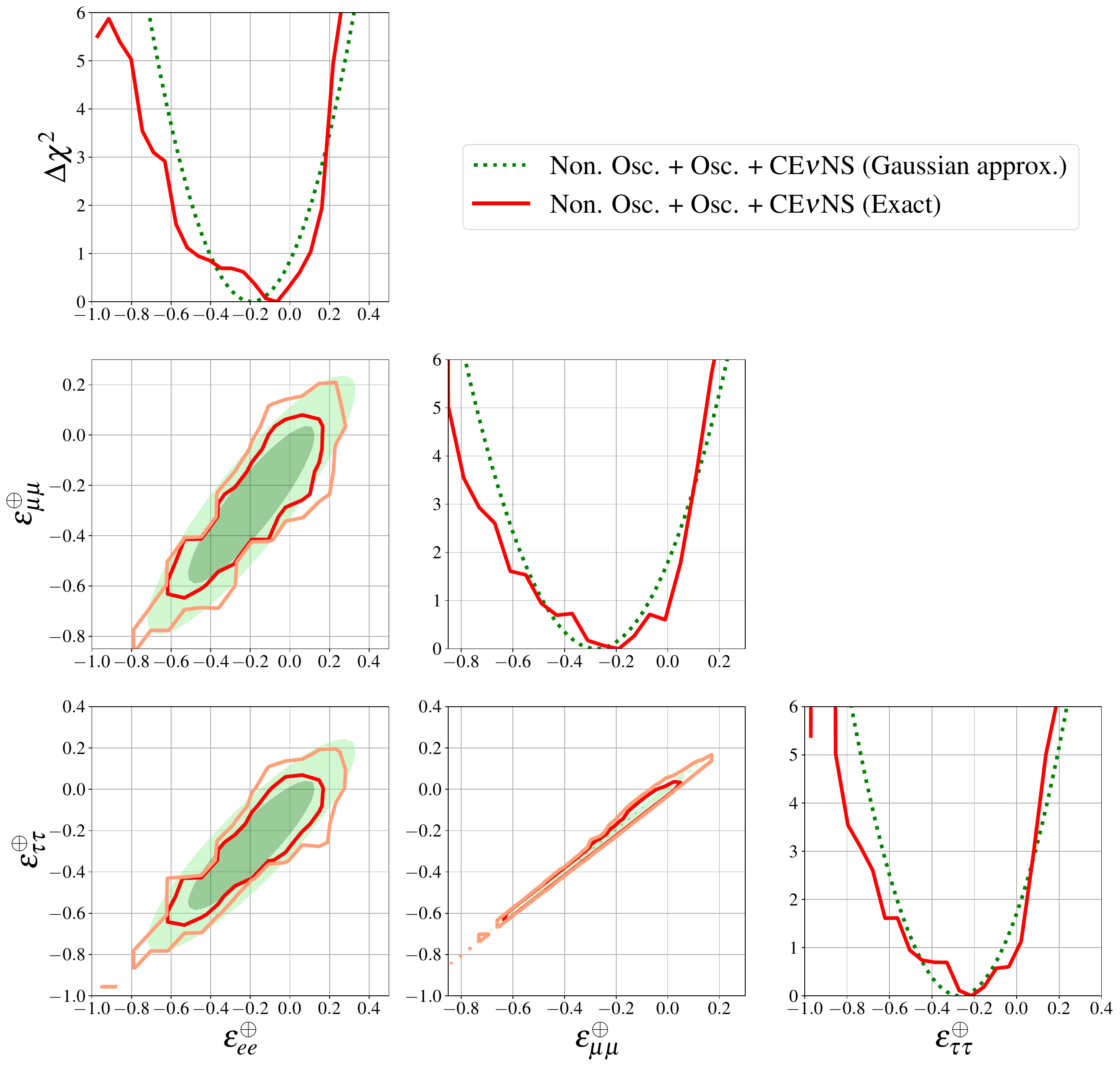}
    \caption{Comparison of considering the Gaussian approximation for the oscillation and CE$\nu$NS data (in green), with respect to using the exact non-Gaussian $\Delta \chi^2$ (in red). The combination with the analysis of non-oscillation data, which is Gaussian, smooths the final results, so the Gaussian approximation leads to a reasonably good agreement.}
    \label{fig:realVSgaussian}
\end{figure}

In Figure~\ref{fig:realVSgaussian} we compare the $\Delta\chi^2$ profiles and contours of the combination of neutrino oscillation and CE$\nu$NS data with low-energy non-oscillation observables, and a gaussian approximation of said profiles and contours. The red lines correspond to the direct combination of the exact fit results of Ref.~\cite{Coloma:2023ixt} for the NSI coefficients $\varepsilon_{\alpha\alpha}^\oplus$ with the corresponding Gaussian SMEFT likelihood from Ref.~\cite{Breso-Pla:2023tnz}, where for the latter we have changed from the basis of SMEFT Wilson coefficients to the basis of $\varepsilon_{\alpha\alpha}^\oplus$ using the matching conditions of Eqs.~\eqref{eq:vector_nsi_matching_eV}-\eqref{eq:vector_nsi_matching_dV}. Alternatively, the green profiles and contours are extracted from first fitting the results on $\varepsilon_{\alpha\alpha}^\oplus$ of Ref.~\cite{Coloma:2023ixt} to a gaussian and then simply combining with the Gaussian SMEFT likelihood. We conclude that, after the combination with the Gaussian SMEFT likelihood, there is reasonable agreement between the exact combined fit and the Gaussian approximation, as shown in Fig.~\ref{fig:realVSgaussian}.

\bibliographystyle{JHEP} 
\bibliography{biblio}

\end{document}